\documentclass[a4paper,fleqn,usenatbib]{mnras}

\usepackage{newtxtext,newtxmath}

\usepackage[T1]{fontenc}
\usepackage{ae,aecompl}

\usepackage{graphicx}	
\usepackage{amsmath}	
\usepackage{amssymb}	

\title[Solar system analogs I]{ Formation of solar system analogs I:\\
  looking for initial conditions through a population synthesis analysis  
 }

\author[M. P. Ronco et al.]{
M. P. Ronco,$^{1,2}$\thanks{E-mail: mpronco@fcaglp.unlp.edu.ar}
O. M. Guilera,$^{1,2}$
G. C. de El\'{\i}a$^{1,2}$
\\
$^{1}$Instituto de Astrof\'{\i}sica de La Plata, CCT La Plata - CONICET, UNLP, Paseo del Bosque S/N, (1900) La Plata, Argentina\\
$^{2}$Facultad de Ciencias Astron\'omicas y Geof\'{\i}sicas, Universidad Nacional de La Plata, Paseo del Bosque S/N (1900), La Plata, Argentina\\
}

\date{Accepted XXX. Received YYY; in original form ZZZ}

\pubyear{2017}

\begin{document}
\label{firstpage}
\pagerange{\pageref{firstpage}--\pageref{lastpage}}
\maketitle


\begin{abstract}
Population synthesis models of planetary systems developed during the last $\sim$15
years could reproduce several of the observables of the exoplanet population, and
also allowed to constrain planetary formation models. We present our planet  
formation model, which calculates the evolution of a planetary system during the gaseous phase. 
The code incorporates relevant physical phenomena for the formation of a planetary system, 
like photoevaporation, planet migration, gas accretion,
water delivery in embryos and planetesimals, a detailed study of the orbital evolution
of the planetesimal population, and the treatment of the fusion between embryos, considering 
their atmospheres. The main goal of this work, unlike other works of planetary
population synthesis, is to find suitable scenarios and physical parameters of the disc 
to form solar system analogs.
We are specially interested in the final planet distributions, and in the final surface
density, eccentricity and inclination profiles for the planetesimal population.
These final distributions will be used as initial conditions for N-body simulations, to
study the post-oligarchic formation in a second work. We then consider different
formation scenarios, with different planetesimal sizes and different type I migration
rates. We find that solar system analogs are favored in massive discs,
with low type I migration rates, and small planetesimal sizes. Besides, those rocky planets
within their habitables zones are dry when discs dissipate.
At last, the final configurations of solar system analogs include information about the mass
and semimajor-axis of the planets, water contents, and the properties of the planetesimal remnants.  
\end{abstract}




\section{Introduction}
\label{sec:sec1}

For a long time, the study of planetary systems was restricted to our own solar system. However, this situation had a drastic change since the discovery of the first exoplanet orbiting a sun-like star, 51 Peg b, in 1995 \citep{MayorQueloz1995}. 51 Peg b, named as Dimidium by the International Astronomical Union in December 2015, is what we know now as a hot-Jupiter, a gas giant planet orbiting very close to its parent star. Since this discovery, the observational field of extrasolar planet search
had an explosion, leading to numerous additional discoveries of planets orbiting other stars. Up to date, 3610 exoplanets that show a wide range of masses, orbits and compositional properties have been discovered orbiting near stars (\url{http://exoplanet.eu/}). Some of them are hot and warm Jupiters, Jupiter-analogs, giant planets at wide orbits, hot-Neptunes, super-Earths, etc. Moreover, much of these planets form part of about 610 multiple planet systems. These systems represent the final stage of a series of complex processes, where a protoplanetary disc evolves into a planetary system, with a few planets and probably, reservoirs of small bodies, like in our solar system. These discoveries also triggered theoretical studies about the formation and evolution of planetary systems. During more than a decade, several models of planet formation have been developed to study the formation of planetary systems and population synthesis with the aim to reproduce the main properties of the observational sample of exoplanets, and to get a better understanding of the main processes of planetary formation \citep[see][for a detailed review]{Benz2014}. 

The first models of planetary population synthesis were developed in the pioneer works of \citet{IdaLin2004a, IdaLin2004b, IdaLin2005,  IdaLin2008a, IdaLin2008b}. In these works, the authors could numerically reproduce several of the main observed properties of the populations of exoplanets, specially the mass versus semimajor-axis diagram (M-a diagram). \citet{Thommes2008} studied the formation of planetary systems with an hybrid model, including N-body interaction between embryos, with the aim to link a mature planetary system with the properties of the protoplanetary disc where it was born. They found that solar system analogs are not common. These systems can form when the timescale for the formation of the giant planets is similar to the timescale of the disc dissipation, undergoing only modest type I migration, and for massive discs. Otherwise, \citet{Miguel2011} found that solar system analogs might not be so rare in the solar neighbourhood, where the formation of such systems could be favoured by massive discs, and where there is not a large accumulation of solids in the inner region. In other case, very slow type I migration is needed to form giant planets. However, \citet{Miguel2011} did not analyze the post-oligarchic growth and long-term evolution of the planetary systems and adopted a different definition for solar system analogs. More recently, \citet{Alibert2013} introduced a distribution of embryos calculating the gravitational interactions between them in their planetary population synthesis model \citep{Alibert2005, Mordasini2009, Fortier2013}, which previously calculates planetary population synthesis but considering only one planet per disc. These authors also showed that the main observed properties of the exoplanets M-a diagram can be reproduced, focusing in low-mass planets. Then, \citet{Pfyffer2015} used the results of \citet{Alibert2013} as initial conditions to calculate the long-term evolution of the planetary population synthesis but only considering the distribution of embryos and without including the remnant of planetesimal distributions. They found that the M-a diagram does not experience significantly changes due to the long-term evolution of the systems, but that the eccentricity distributions of such systems do not match the observed ones in the known exoplanets. Moreover, \citet{Ida-et-al-2013} also introduced gravitational interactions between embryos in their planetary population synthesis model using a Monte Carlo approach (not calculating N-body integrations) founding that the observed planetary mass-eccentricity and M-a diagrams can be reproduced. However, in most of the cases, the evolution of the planetesimal population is treated in a simplified way and is not taken into account in the calculations of the long-term evolution of the systems. 

On the other hand, several works study the formation of terrestrial planets in the post-oligarchic growth regime and the water delivery via purely N-body calculations considering a distribution of embryos and planetesimals after the dissipation of the primordial nebula \citep{Chambers2001, Raymond2004, Raymond2005, Raymond2006, Raymond2009, Obrien2006, Obrien2014, Walsh2011, RoncodeElia2014}. In general, the orbital initial conditions for the distributions of embryos and planetesimals are typically selected \emph{ad hoc}. However, \citet{Ronco2015} showed that more realistic initial conditions lead to different accretion histories of the planets that remain in the habitable zone compared to the case of ad hoc initial conditions. 
     
The aim of this first work is to improve our model of planet formation during the gaseous phase to obtain accurate initial conditions for the distributions of planets and planetesimals, to perform future high resolution N-body calculations for the analysis of the formation of terrestrial planets and water delivery in solar system analogs. To do so, we first include the most relevant physical phenomena that take place in the formation of a planetary system during the gaseous phase. Then, we perform a planetary population synthesis calculation in order to obtain the main properties of the protoplanetary discs that lead to the formation of solar system analogs. This work is then organized as follow: in Sec.~\ref{sec:sec2} we describe our planet formation model; then, in Sec.~\ref{sec:sec12} we describe the initial conditions for the development of
our planetary systems, in Sec.~\ref{sec:sec15} we show the results, and in Sec.~\ref{sec:sec21}, we discuss the conclusions.


\section{Description of our model of planet formation}
\label{sec:sec2}

We have improved a model of planet formation based on previous
works \citep{Guilera2010, Guilera2011, Guilera2014},
which we now called  P{\scriptsize LANETA}LP. This code describes the evolution in time of a planetary system
during the gaseous phase and incorporates many of the most important physical phenomena for the formation of planetary systems. P{\scriptsize LANETA}LP is a code developed
almost entirely in FORTRAN 95/2003 and has the advantage of being programmed in a modular way so as to be able to \emph{turn on} or \emph{turn off}
the different physical phenomena according to the user's choice.

Our model presents a protoplanetary disc, which is characterized by two components: a gaseous component, evolving due to an
$\alpha$-viscosity driven accretion and photoevaporation processes, and a solid component represented by a planetesimal population being 
subject to accretion and ejection by the embryos, and radial drift due to gas drag. Besides, the model presents an embryo population which is also part
of the solid component of the disc and which grows by accretion of planetesimals, gas,
and due to the fusion between them taking into account their atmospheres. The new improvements made in P{\scriptsize LANETA}LP with respect 
to previous versions of this code \citep{Guilera2010, Guilera2011, Guilera2014} are described in detail in the next subsections.

\subsection{Protoplanetary disc structure}
\label{subsec:subsec3}

The gaseous component is characterized by the gas surface density $\Sigma_{\text{g}}(R)$, where $R$ is the radial coordinate in the midplane
of the disc, and the planetesimal population is characterized by the planetesimal surface density $\Sigma_{\text{p}}(R)$. Although our model 
allows to include a discrete 
planetesimal size distribution described by $\Sigma_{\text{p}}(R,r_{\text{p}})$, where $r_{\text{p}}$ represents the different sizes of the 
planetesimals, we will consider in this work only one size of planetesimals for each simulation. 
Both $\Sigma_{\text{g}}(R)$ and $\Sigma_{\text{p}}(R)$ determine the mass distribution along the 
protoplanetary disc. In this work, as in previous ones \citep{deElia2013,Ronco2015,Dugaro2016} we consider the 
surface density profile suggested by \citet{Andrews2010}, who found that the discs observed in the Ophiuchus star-forming region can 
be represented by

\begin{equation}
\Sigma_{\text{g}}(R) = \Sigma^{0}_{\text{g}}\left(\dfrac{R}{R_{\text{c}}}\right)^{-\gamma}e^{-\left(\frac{R}{R_{\text{c}}}\right)^{2-\gamma}},
\label{eq:gas}
\end{equation}
where $R_{\text{c}}$ is a characteristic radius of the disc, $\gamma$ is the exponent that represents the surface density gradient and 
$\Sigma_{\text{g}}^{0}$ is a normalization constant, which is a function of the mass of the disc $M_{\text{d}}$, $R_{\text{c}}$ and $\gamma$, 
that can be obtained by integrating Eq. \ref{eq:gas} over the total area of the disc. Assuming that the metallicity
along the disc is the same as that of the central star, and considering that the dust sediments and coagulates very quickly to form a 
mid-plane planetesimal disc, is that the planetesimal surface density profile is given by
\begin{equation}
\Sigma_{\text{p}}(R) = \Sigma^{0}_{\text{p}}\eta_{\text{ice}}\left(\dfrac{R}{R_{\text{c}}}\right)^{-\gamma}e^{-\left(\frac{R}{R_{\text{c}}}\right)^{2-\gamma}},
\label{eq:solids}
\end{equation}
where  $\Sigma^{0}_{\text{p}} = z_0\Sigma^{0}_{\text{g}}10^{[\text{Fe/H}]}$ is a normalization constant, $z_0 = 0.0153$ is the primordial
abundance of heavy elements in the Sun \citep{Lodders2009}, $[\text{Fe/H}]$ is the stellar metallicity, and $\eta_{\text{ice}}$ is a
parameter that represents an increase in the amount of solid material due to the condensation of water beyond the snow line, given by
\begin{equation}
  \eta_{\text{ice}}= 
  \begin{cases}
    1 & \text{ if $R \ge R_{\text{ice}}$},  \\
    \\
    {\dfrac{1}{\beta}} & \text{ if $R < R_{\text{ice}}$},
  \end{cases} 
\label{eq:eta}
\end{equation}
where $R_{\text{ice}}$ represents the position in the disc where water condensates, usually at $170^{\circ}$~K,  
and the factor $\beta$ represents the amount of material 
condensed. In our solar system, $\beta$ usually takes a value of approximately 2 \citep{Lodders2003,Lodders2009}.

The values of the free parameters involved in the structure of the protoplanetary disc, like the mass of the disc $M_{\text{d}}$, the 
characteristic radius $R_{\text{c}}$, the surface density gradient $\gamma$, and the factor $\beta$ will be discussed in the next section.

\subsubsection{Evolution of the gaseous component}
\label{subsubsec:subsubsec4}

As it was mentioned before, 
we improved the evolution of the gaseous component considering a viscous accretion
disc \citep{Pringle1981} with photoevaporation.  

The evolution in time of the gas surface
density is then represented by a diffusion equation 
\begin{equation}
  \frac{\partial \Sigma_{\text{g}}}{\partial t}= \frac{3}{R}\frac{\partial}{\partial R} \left[ R^{1/2} \frac{\partial}{\partial R} \left( \nu \Sigma_{\text{g}} R^{1/2}  \right) \right] + \dot{\Sigma}_{\text{w}}(R) , 
\label{eq:gas-evolution}
\end{equation}
where $\nu= \alpha c_{\text{s}} H_{\text{g}} $ represents the Shakura and Sunyaev viscosity \citep{Shakura1973} with $\alpha$ a
constant a-dimensional parameter along the disc, $c_{\text{s}}$ the sound speed, $H_{\text{g}}$ the height scale of the disc, and $\dot{\Sigma}_{\text{w}}(R)$
the photoevaporation sink term. The sound speed is given by 
\begin{equation}
  c_{\text{s}}= \sqrt{ \frac{\gamma_{\text{g}} k_{\text{B}} \text{T}}{\mu_{\text{H}_2} m_{\text{H}_2}} },
  \label{eq:eq2-sec2-2}
\end{equation}  
where $\gamma_{\text{g}}= 5/3$, $k_{\text{B}}$ is the Boltzmann constant, and $\mu_{\text{H}_2}$ and $m_{\text{H}_2}$ are the molecular weight and mass of molecular
hydrogen, respectively. Following \citet{Hayashi1981} and \citet{IdaLin2004a}, we consider a temperature profile given by 
\begin{equation}
  \text{T}= 280 \left( \frac{R}{1~\text{au}} \right)^{-1/2} ~ \text{K},
  \label{eq:eq3-sec2-2}
\end{equation}
where, according to this, $R_{\text{ice}}$ is defined as 2.7~au. Finally,  
the height scale of the disc is $H_{\text{g}} = \sqrt{2}c_{\text{s}}/\Omega_{\text{k}}$ where $\Omega_{\text{k}}$ is the keplerian frequency. It is important to 
note that we are considering a flare disc, where $H_{\text{g}} \propto R^{5/4}$. 

Finally, $\dot{\Sigma}_{\text{w}}(R)$ represents the photoevaporation rate. There are different photoevaporation regimes, and they depend 
on which is the source 
that emits irradiation. This source could be the central star \citep{Dullemond2007,DangeloMarzari2012}, or external stars that irradiate 
the environment in where the disc forms \citep{VerasArmitage2004}. If our sun was formed in a densely populated environment, 
the high stellar density could lead to encounters with stars that could affect and modify the size of the disc by photoevaporation \citep[see][and references therein]{MoroMartin2008}. However, since up to date the environment 
in which our solar system was formed is still under discussion, and due to the fact that photoevaporation by the central star is, probably, 
the most important photoevaporation regime that affects the planet formation region, we only consider this case. Thus, 
following the model of \citet{Dullemond2007}, also adopted by \citet{DangeloMarzari2012}, $\dot{\Sigma}_{\text{w}}(R)$ is given by
\begin{equation}
  \dot{\Sigma}_{\text{w}}(R) = 
  \begin{cases}
    \dot{\Sigma}^{g}_{\text{w}}\exp{\left[\dfrac{1}{2}\left(1-\dfrac{R_{\text{g}}}{R}\right)\right]}\left(\dfrac{R_{\text{g}}}{R}\right)^2 & \text{ if $R \le R_{\text{g}}$},  \\
    \\
    \dot{\Sigma}^{g}_{\text{w}}\left(\dfrac{R_{\text{g}}}{R}\right)^{5/2} & \text{ if $R > R_{\text{g}}$},
  \end{cases} 
\label{eq:eta2}
\end{equation}
where $R_{\text{g}}$ is the radius of the gas surface layer at which the sound speed equals the local keplerian velocity, and beyond which the gas is
disengaged from the disc. The photoevaporation rate at $R_{\text{g}}$ is represented by $\dot{\Sigma}^{g}_{\text{w}}$ and is given by
\begin{equation}
  \dot{\Sigma}^{g}_{\text{w}} = 1.16 \times 10^{-11}f^{0.5}_{41}\left(\dfrac{1\text{au}}{R_{\text{g}}}\right)^{1.5}\dfrac{\text{M}_\odot}{\text{au}^{2}\text{yr}},
  \label{eq:eq4-sec2-2}
\end{equation}
where $f_{41}$ is the rate of EUV ionizing photons emitted by the star in units of $10^{41}$s$^{-1}$. It is worth noting that $R_{\text{g}}$ is usually
taken to be 10~au for a solar-mass star.

Equation \ref{eq:gas-evolution} is solved on a disc defined between 0.1~au and 1000~au, using 5000 radial bins logarithmically equally spaced. It is worth remarking that we consider that the disc is dissipated when the mass of gas becomes lower than $10^{-6}~M_{\odot}$.

\subsubsection{Evolution of the planetesimal population}
\label{subsubsec:subsubsec5}

In our model, the planetesimal population evolves by the drift of planetesimals due to the nebular gas including the 
Epstein, Stokes and quadratic regimes, and the accretion and ejection by the embryos. We also consider that the evolution of the 
eccentricities and inclinations of the planetesimals 
are due to two principal processes: the embryo gravitational excitation \citep{Ohtsuki2002}, and the damping due to the nebular gas 
drag \citep{Rafikov2004,Chambers2008}. The treatment of these processes is described in detail in \citet{Guilera2014}. Then, 
the evolution in time of the planetesimal surface density is modeled with a continuity equation given by 

\begin{equation}
  \frac{\partial \Sigma_{\text{p}}(R)}{\partial t} - \frac{1}{R} \frac{\partial}{\partial R} \left[ Rv_{\text{mig}}(R)\Sigma_{\text{p}}(R) \right] = \dot{\Sigma}^{\text{tot}}_{\text{p}}(R), \label{eq:densi-planetes-1} 
\end{equation}
\begin{equation}
\dot{\Sigma}^{\text{tot}}_{\text{p}}(R) = \dot{\Sigma}^{\text{ac}}_{\text{p}}(R) + \dot{\Sigma}^{\text{sc}}_{\text{p}}(R) + \dot{\Sigma}^{\text{ej}}_{\text{p}}(R), 
\label{eq:densi-planetes-2}
\end{equation}
where $v_{\text{mig}}$ is the planetesimal migration velocity and $\dot{\Sigma}^{\text{ac}}_{\text{p}}(R)$, $\dot{\Sigma}^{\text{sc}}_{\text{p}}(R)$, and $\dot{\Sigma}^{\text{ej}}_{\text{p}}(R)$ represent 
the sink terms due to the accretion by the embryos, due to the probability of scattered planetesimals, and due to the ejection of
planetesimals with high eccentricities, respectively.
In this work, we incorporate two different ejection rates of planetesimals that play different roles. It is important
to remark that the ejected planetesimals are directly removed from the planetary system and are not relocated in a different
orbit.
The first one, $\dot{\Sigma}^{\text{sc}}_{\text{p}}(R)$, considers the ejection rate of planetesimals proposed by 
\citet{IdaLin2004a} and also adopted by \citet{Alibert2005}, that takes into account the probability of planetesimal scattering 
rather than collisions during close encounters, and is 
\begin{equation}
\dot{\Sigma}^{\text{sc}}_{\text{p}}(R) = \dfrac{1}{4}\left(\dfrac{a_{\text{P}}M_{\text{P}}}{M_\star\tilde{R}_{\text{C}}}\right)^{2}\dot{\Sigma}^{\text{ac}}_{\text{p}}(R),
\label{eq:ejection_rate}
\end{equation}
where $a_{\text{P}}$ and $M_{\text{P}}$ are the semimajor-axis and the total mass of the planet, respectively, and $\tilde{R}_{\text{C}}$ is the 
planet's enhanced radius (see next section). 
The second one, $\dot{\Sigma}^{\text{ej}}_{\text{p}}(R)$, considers the ejection of planetesimals that reach values of 
the eccentricity higher than $0.99$. As the gas disc evolves, the gas surface density decreases, and because of this, the eccentricities 
of the planetesimals that are in the vicinity of the massive planets are no longer efficiently damped by the gas, and therefore, 
they increase considerably. At this time,  $\dot{\Sigma}^{\text{ej}}_{\text{p}}(R)$ becomes effective. As in \citet{Laakso2006}, we consider
that the probability of a particle to be ejected per unit time is constant. This implies that 
$\dot{\Sigma}^{\text{ej}}_{\text{p}}= -\rho\Sigma_{\text{p}}$, where we adopt \emph{ad hoc} values of $\rho$ between 0.001 yr$^{-1}$ and 0.01 yr$^{-1}$. These values
represent the ejection of the 0.1\% and 1\%, respectively, of the available planetesimals in each radial bin around the planet at
each time-step, finding no differences in the final results. 

We use a full implicit upwind-downwind mix method to solve Eq.~\ref{eq:densi-planetes-1}, considering density equal to zero as boundary conditions.
We do not allow changes greater than 10\% for the planetesimal surface density profile in each timestep and in each radial bin.

\subsection{Growth and orbital evolution of the protoplanetary embryos}
\label{subsec:subsec6}

At the beginning of our model, an embryo population is immersed in the protoplanetary disc. Here, we describe
 the details on the embryos growth and orbital evolution.

\subsubsection{Solid accretion by the embryos}
\label{subsubsec:subsubsec7}

The embryos grow within the oligarchic regime, and the accretion of planetesimals is well described 
by the \emph{particle in a box} approximation \citep{Inaba2001}
\begin{equation}
  \dfrac{\text{d}M_{\text{C}}}{\text{d}t} = \dfrac{2\pi\Sigma_{\text{p}}({a_{\text{P}}})R^{2}_{\text{H}}}{P}P_{\text{coll}}.
  \label{eq:solidaccretionrate}
\end{equation}
In this equation $M_{\text{C}}$ is the mass of the core, $\Sigma_{\text{p}}({a_{\text{P}}})$ is the planetesimal surface density at the planet location, $R_{\text{H}}$ is the Hill radius, $P$ is the orbital period, and $P_{\text{coll}}$ is the collision probability. $P_{\text{coll}}$ is a function of the enhanced radius $\tilde{R}_{\text{C}}$, the Hill radius of the planet, and the relative velocity of planetesimals, $P_{\text{coll}} = P_{\text{coll}}(\tilde{R}_{\text{C}},R_{\text{H}},v_{\text{rel}})$. The fact that $P_{\text{coll}}$ is function of the enhanced radius $\tilde{R}_{\text{C}}$ instead of the core radius $R_{\text{C}}$ is due to the fact that we are considering the drag force experienced by planetesimals when entering the planetary envelope. Here, $v_{\text{rel}}= \sqrt{5e^{2}/8+i^{2}/2}~v_{\text{k}}$, where $e$ represents the eccentricity, $i$ the inclination, and $v_{\text{k}}$ the keplerian velocity of the planetesimal population (see \citet{Guilera2010} for further details). It is worth remarking that the
embryos evolve in circular and coplanar orbits during all the simulation. 

In contrast to our previous works, we do not solve the equations of transport and structure for the envelope, thus we calculate $\tilde{R}_{\text{C}}$ following the works of \citet{OrmelKobayashi2012} and \citet{Chambers2014} as
\begin{equation}
  \tilde{R}_{\text{C}}= 
  \begin{cases}
    \dfrac{R_\text{B}}{1 + \frac{2W(\sigma_{\text{crit}}-1) + \ln(\sigma_{\text{crit}})}{\gamma_{\text{g}}}},~\text{if}~\sigma_{\text{crit}} < \sigma_t, \\
    \\
    \dfrac{R_\text{B}}{\frac{R_\text{B}}{R_t} + \frac{4}{\gamma_{\text{g}}} (4W)^{1/3}(\sigma_{\text{crit}}^{1/3} - \sigma_t^{1/3})},~\text{if}~\sigma_{\text{crit}} > \sigma_t,
  \end{cases}
   \label{eq:enhanced-radius}
\end{equation}
where 
\begin{align}
  \dfrac{R_\text{B}}{R_t} &= 1 + \dfrac{2W(\sigma_t - 1) + \ln(\sigma_t)}{\gamma_{\text{g}}}, \\
  W &= \dfrac{3\kappa_e L P_{\text{neb}}}{64 \pi \sigma_B G M_{\text{C}} T_{\text{neb}}^4}, \\
  \sigma_t &= \dfrac{1}{5W}, \\
  \sigma_{\text{crit}} &= \dfrac{\rho_{\text{p}}r_{\text{p}}v_{\text{rel}}^2}{3GM_{\text{C}}\rho_{\text{neb}}}, \\
  L &= \left(\dfrac{GM_{\text{C}}}{R_{\text{C}}}\right) \dfrac{dM_{\text{C}}}{dt}, \\
  R_\text{B} &= \dfrac{GM_{\text{C}}}{c_{\text{s}}^2}.
 \label{eq:enhanced-radius-auxiliar}
\end{align}
In the previous equations, $R_{\text{B}}$ is the Bondi radius of the planet, $L$ the luminosity generated due to the accretion of planetesimals, $G$ the universal gravitational constant, $R_{\text{C}}$ the radius of the core, $\rho_{\text{p}}$ the planetesimal density, $\sigma_B$ the Stefan-Boltzmann constant, $\kappa_e= 0.01~\text{cm}^2 \text{g}^{-1}$ the envelope opacity, and $P_{\text{neb}}, T_{\text{neb}}, \rho_{\text{neb}}$ the local pressure, temperature and density of the nebular gas, respectively.     

\subsubsection{Gas accretion by the embryos}
\label{subsubsec:subsubsec8}

As the embryos grow, they are able to retain a gaseous envelope that is capable to maintain hydrostatic equilibrium. 
Initially, the gas accretion rate is much lower than the planetesimal accretion rate, thus, the core of the embryo grows faster 
than the corresponding envelope. However, when the core of the planet reaches a critical mass, the gas accretion begins to be substantial. 
Following \citet{IdaLin2004a} the critical mass is given by 
\begin{equation}
  M_{\text{crit}} \sim 10\left(\dfrac{\dot{M}_{\text{C}}}{10^{-6}\text{M}_\oplus \text{yr}^{-1}}\right)^{0.25}\text{M}_\oplus,
  \label{eq:critical_mass}
\end{equation}
where, as we have already seen, $\dot{M}_{\text{C}}$ is the planetesimal accretion rate.
The gas accretion rate can be estimated by
\begin{equation}
 \dot{M}_{\text{KH}} = \dfrac{\text{d}M_{\text{g}}}{\text{d}t} = \dfrac{M_{\text{P}}}{\tau_{\text{g}}} 
  \label{eq:gas_accretion_rate}
\end{equation}
where $M_{\text{P}}$ is the total mass of the planet and $\tau_{\text{g}}$ is the characteristic 
Kelvin-Helmholtz growth timescale of the envelope. According to the results of the giant planet formation model of 
\citet{Guilera2010,Guilera2014}, and following the prescription of \citet{IdaLin2004a}, $\tau_{\text{g}}$ is fitted by
\begin{equation}
 \tau_{\text{g}} = 8.35\times10^{10}(M_{\text{P}}/\text{M}_\oplus)^{-3.65}\text{yr}.
 \label{eq:KH time_scale}
\end{equation}

Note that, unlike \citet{Miguel2011}, we found a different exponent in the dependence with the mass of the planet of 
$\tau_{\text{g}}$. These authors found an exponent of $-4.89$. This difference is due to the fact that the fit on the gas 
accretion has been done 
for slightly different giant planet formation models\footnote{\citet{Miguel2011} used the results of \citet{Fortier2009}.}.

It is important to remark that $\dot{M}_{\text{KH}}$
is valid as long as it is less than the maximum rate at which the gas can be delivered by the disc onto the planet, which, 
following \citep{Mordasini2009}, is
\begin{equation}
 \dot{M}_{\text{disc}} = \dfrac{\text{d}M_{\text{g}}}{\text{d}t} = 3\pi\nu\Sigma_{\text{g}}(R).
 \label{eq:disc_accretion_rate}
\end{equation}
Thus, we consider that the effective gas accretion rate is
\begin{equation}
 \dfrac{\text{d}M_{\text{g}}}{\text{d}t} = \text{min}[\dot{M}_{\text{KH}},\dot{M}_{\text{disc}}]. 
 \label{eq:KH time_scale_2}
\end{equation}

The process of gas accretion by the planet is also limited if the planet is able to open a gap.
Through high resolution hidrodynamical simulations, \citet{TanigawaIkoma2007} found that the gas
accretion rate decreases significantly when a planet opens a gap in what they called the ``gap-limiting'' case. 
Following this idea, \citet{TanigawaIkoma2007}
developed an analytic formula for the accretion rate also adopted by \citet{XiaoJin2015}

\begin{equation}
  \dot{M}_{\text{GP}} = \dfrac{\text{d}M_{\text{g}}}{\text{d}t} = \dot{A}\Sigma_{\text{acc}}, 
  \label{eq:gap_limiting_case}
\end{equation}
where
\begin{equation}
  \dot{A} = 0.29\left(\dfrac{H_{\text{g,P}}}{a_{\text{P}}}\right)^{-2}\left(\dfrac{M_{\text{P}}}{M_{\star}}\right)^{4/3}a^{2}_{\text{P}}\Omega_{\text{k,P}} 
  \label{eq:gap_limiting_case_2}
\end{equation}
with $\Sigma_{\text{acc}} = \Sigma(x=2R_{\text{H}})$ and,

\begin{equation}
  \Sigma(x) =
  \begin{cases}
    \Sigma_{\text{g}}\text{exp}\left[-\left( \dfrac{x}{l} \right)^{-3}\right] & \text{ if $x > x_{\text{m}}$}, \\
    \\
    \!\begin{aligned}
      & \Sigma_{\text{g}}\text{exp}\Bigg[-\dfrac{1}{2}\left(\dfrac{x}{H_{\text{g,P}}}-\dfrac{5}{4}\dfrac{x_{\text{m}}}{H_{\text{g,P}}}\right)^{2} \\
    \\
    & + \dfrac{1}{32}\left(\dfrac{x_{\text{m}}}{H_{\text{g,P}}}\right)^{2} - \left(\dfrac{x_{\text{m}}}{l}\right)^{-3}\Bigg] \\
    \end{aligned} & \text{if $x \le x_{\text{m}}$}, 
  \end{cases}
\label{eq:cases}
\end{equation}

and $l$ and $x_{\text{m}}$ are defined as

\begin{equation}
  l = 0.146\left(\dfrac{\nu}{10^{-5}a^{2}_{\text{P}}\Omega_{\text{k,P}}}\right)^{-1/3}\left(\dfrac{M_{\text{P}}}{10^{-3}M_{\star}}\right)^{2/3}a_{\text{P}},
  \label{eq:cases1}
\end{equation}
\begin{equation}
  x_{\text{m}} = 0.207\left(\dfrac{H_{\text{g}}}{0.1a_{\text{P}}}\dfrac{M_{\text{P}}}{10^{-3}M_{\star}}\right)^{2/5}\left(\dfrac{\nu}{10^{-5}a^{2}_{\text{P}}\Omega_{\text{k,P}}}\right)^{-1/5}a_{\text{P}},
  \label{eq:cases2}
\end{equation}
where  $M_{\star}$ is the mass of the star, and the sub-index $\text{P}$ represents at the location of the planet.
Therefore, the gas accretion rate adopted to limit the gas accretion after the planet opens a gap is defined as
\begin{equation}
 \dfrac{\text{d}M_{\text{g}}}{\text{d}t} = \text{min}[\dot{M}_{\text{KH}},\dot{M}_{\text{disc}},\dot{M}_{\text{GP}}]. 
 \label{eq:cases3}
\end{equation}

\begin{figure}
  \includegraphics[angle= 270, width=\columnwidth]{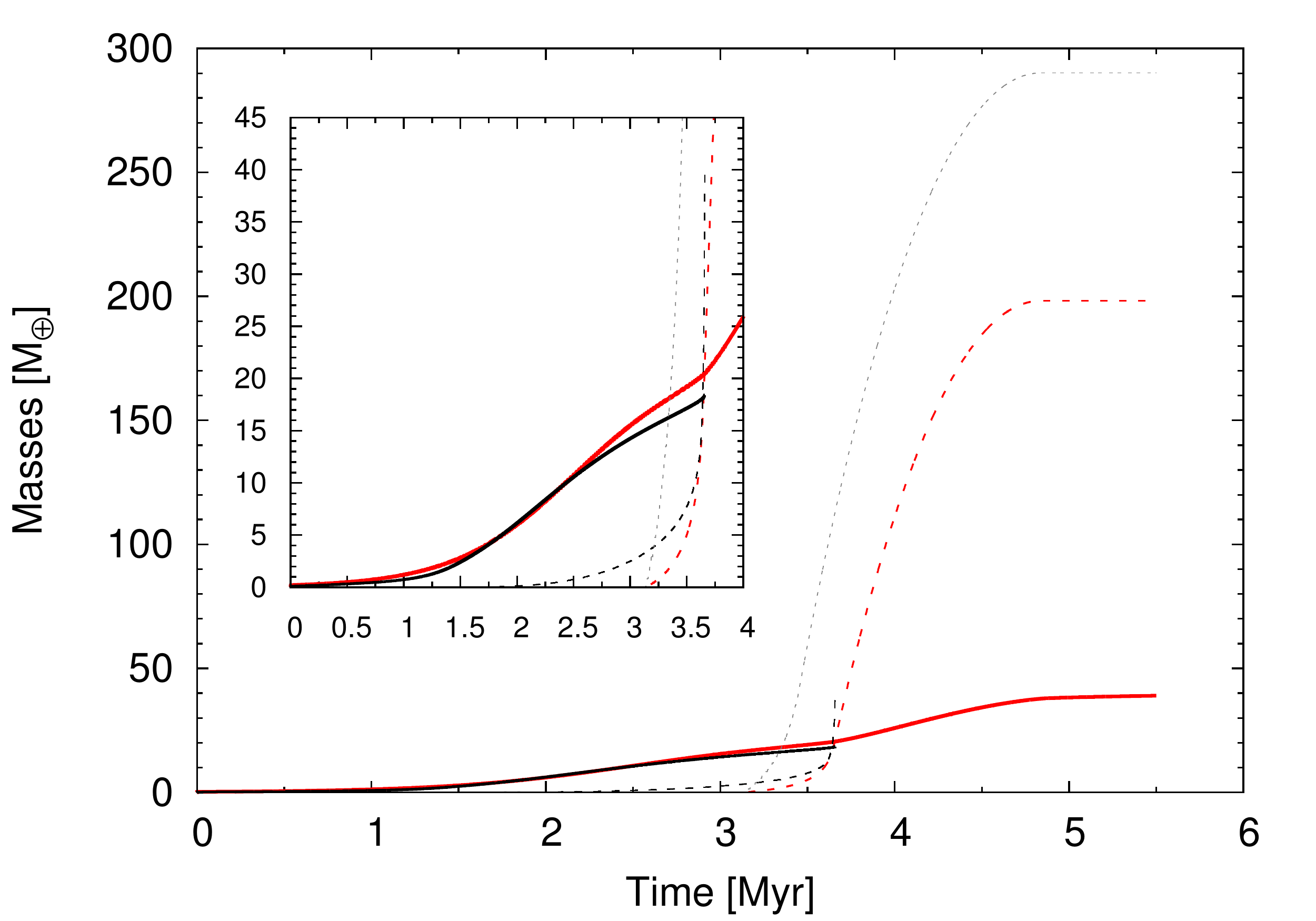}
  \caption{Mass of the core (solid line) and mass of the envelope (dashed line) as a function of the time for the in situ formation of a planet at 5~au. The black lines correspond to the giant planet formation model developed by \citet{Guilera2010, Guilera2014}, while the red lines correspond to P{\scriptsize LANETA}LP. The grey line corresponds to a model using the gas accretion prescription given by \citet{Miguel2011}. This simulation has been developed for a disc with a mass of $0.1\text{M}_\odot$ using $\gamma= 1$, $R_{\text{c}}= 25$~au, $\alpha= 10^{-4}$, and using planetesimals of 10~km. The small square in the figure represents a zoom of the moment at which the gas accretion is triggered.}
    \label{fig:fig1}
\end{figure}

As an example, in Fig.~\ref{fig:fig1} we plot a comparison for the in situ formation of a giant planet at 5~au between the giant planet formation model \citep{Guilera2010, Guilera2014} and P{\scriptsize LANETA}LP. For the first case, when the mass of the envelope reaches $\sim 40\text{M}_{\oplus}$, the giant planet formation model of \citet{Guilera2010, Guilera2014} is not able to solve any more the equations of transport and structure for the envelope and the simulation stops. It is important to note that there is no limitation in the gas accretion for the first case. We can see that P{\scriptsize LANETA}LP reproduces well the growth of the envelope, specially when the planet begins the gaseous runaway growth (see the zoom in Fig.~\ref{fig:fig1}). We also plot the growth of the envelope using the prescription of \citet{Miguel2011} for the gas accretion rate. Different gas accretion rates lead to different final masses of the planet. The small difference between the value of the mass of the cores for the cross-over mass (when the mass of the envelope equals the mass of the core) is because the enhanced radius is calculated in different ways in each model.

It is important to note that the gas accretion rate onto planets is not considered as a sink term in the time evolution of 
the gas surface density profile. Therefore our model controls that the mass of gas accreted by the planets of the system is not 
greater than 90\% of the mass of gas available in the disc at each time step.

\subsubsection{Fusion of embryos and planets with atmospheres}
\label{subsubsec:subsubsec9}

As we have already mentioned, embryos grow by accretion of gas and planetesimals, but they also grow due to collisions with each other. In the present study, we consider that when the distance between two embryos becomes smaller than 3.5 mutual Hill radii, they merge into one object. From this, it is important to specify the mass of the resulting object, which will depend on the physical properties of the interacting 
bodies. 
In fact, if the embryos that participate in a collision are not very massive, their individual gaseous envelopes are negligible. In this case, we assume perfectly inelastic collisions, from which the mass of the resulting embryo will be the sum of the individual masses of the interacting embryos.  
If one or both of the interacting bodies is sufficiently massive to have a significant gaseous envelope, it is necessary to specify the core and envelope masses of the resulting body. On the one hand, the individual cores are perfectly merged in order to define the final core. On the other hand, the evolution of the gaseous envelopes is computed from the studies developed by \citet{InamdarSchlichting2015}. These authors analyzed the formation of super-Earths and mini-Neptunes and examined how much gaseous envelope could have been accreted by planetary embryos before giant impacts and how much could be retained throughout the giant impact phase. \citet{InamdarSchlichting2015} computed the global atmospheric mass loss fraction $\chi_{\text{loss}}$ for ratios of envelope mass to core mass $M_{\text{E}}/M_{\text{C}}$ between 10$^{-1}$ and 10$^{-6}$ as a function of 
$v_{\text{imp}}m_{\text{imp}}/(v_{\text{esc}}M_{\text{C}})$, being $v_{\text{imp}}m_{\text{imp}}$ the impactor momentum, and $v_{\text{esc}}$ the escape velocity of the target core of mass $M_{\text{C}}$. The values of $\chi_{\text{loss}}$ used in the present work are those exposed in Fig. 5 of \citet{InamdarSchlichting2015}. From these considerations, if bodies $i$ and $j$ collide, the core mass $M_{\text{C}}$ of the resulting object will be given by 
\begin{equation}
M_{\text{C}} = M^{i}_{\text{C}} + M^{j}_{\text{C}},
\end{equation}
where $M^{i}_{\text{C}}$ and $M^{j}_{\text{C}}$ are the core mass of the body $i$ and $j$, respectively, while the final envelope mass $M_{\text{E}}$ will be      
\begin{equation}
M_{\text{E}} = M^{i}_{\text{E}}(1 - \chi_{\text{loss}}^{i}) + M^{j}_{\text{E}}(1 - \chi_{\text{loss}}^{j}),
\end{equation}
where $M^{i}_{\text{E}}$ and $M^{j}_{\text{E}}$ are the envelope mass of the body $i$ and $j$, respectively, and $\chi_{\text{loss}}^{i}$ and $\chi_{\text{loss}}^{j}$ the atmospheric mass loss fraction of the body $i$ and $j$, respectively.

Finally, by conservation of angular momentum, we compute the semimajor-axis of the resulting body by 
\begin{equation}
a = \left(\frac{a^{0.5}_{i}(M^{i}_{\text{C}} + M^{i}_{\text{E}}) + a^{0.5}_{j}(M^{j}_{\text{C}} + M^{j}_{\text{E}})}{M_{\text{C}} + M_{\text{E}}}\right)^{2}, 
\end{equation}
being $a_{i}$ and $a_{j}$ the semimajor-axis of the body $i$ and $j$, respectively.

It is important to remark that the studies of \citet{InamdarSchlichting2015} concerning the loss of atmospheric mass were developed for planets with masses in the range of the Neptunes and the so-called super-Earths and initial envelopes less than 10\% of the core mass. Due to the absence of works about atmospheric mass loss by giant impacts in large atmospheres, we also used the results derived by \citet{InamdarSchlichting2015} to treat collisions that involve giant planets. However, it is important to note that, in our study, the final mass of a giant planet is not sensitive to this consideration. This is due to the fact that a planet that is the final result of a merger between two previous planets, is still growing in the disc. Thus, if this final planet has a significant core mass, it quickly accretes large amounts of gas. In fact, \citet{BroegBenz2012} studied in detail the effect of this situation on the gas accretion rate of a planet. They found that initially, most of the envelope can be ejected, but afterwards, the gas is re-accreted very fast.

\subsubsection{Type I and type II migration}
\label{subsubsec:subsubsec10}

A planet immersed in a protoplanetary disc modifies the local gas surface density and the mutual gravitational interactions between them, which lead to an exchange of angular momentum between the gas and the planet. This produces torques that cause that the planet migrates towards 
or away from the central star. Different types of migration regimes have been studied by many authors during the last two decades
\citep{Ward1997,Tanaka2002,Masset2006,Paardekooper2010,Paardekooper2011}. 
These regimes modify the local density profile and the planets orbits
in different ways depending on the mass of the planet and on the local properties of the disc.

Type I migration \citep{Ward1997} affects those planets that are not massive enough to open a gap in the gaseous disc. In
idealized vertically isothermal discs, type I migration produces always inward migration and very fast migration rates \citep{Tanaka2002}.
Since type I migration is much faster than the gas dissipation timescales of the discs and therefore much faster than the formation
timescales of the cores of giant planets, the survival of the planets, particularly of giant planets, results to be a difficult process to fulfill \citep{Papaloizou2007}.
Although these migration rates have been corroborated by numerical simulations \citep{DAngelo2003,DAngeloLubow2008}, many authors had to reduce the migration rates
by a constant factor in order to reproduce observations \citep{IdaLin2004a,Alibert2005,Mordasini2009,Miguel2011}.

It is important to note that more recent works which consider non isothermal discs \citep{Paardekooper2010,Paardekooper2011} have shown that type I migration
could be outward. Even in isothermal discs, type I migration could produce outward migration if a full magneto-hydrodynamic (MHD) disc turbulence is considered
\citep{Guilet2013}. Moreover, \citet{BenitezLlambay2015} showed that the heating torques produced by a planet while the process of solid accretion is given,
are capable of slowing, halting or even reversing the migration of the planet in the protoplanetary disc if the mass of the planet is lower than $5\text{M}_\oplus$.

In order to be consistent with our model of planet formation, P{\scriptsize LANETA}LP, we use the migration rates for isothermal discs of \citet{Tanaka2002} but considering 
different reduction factors $f_{\text{migI}}$, given by 
\begin{equation}
 \left(\dfrac{\text{d}a_{\text{P}}}{\text{d}t}\right)_{\text{migI}} = -2f_{\text{migI}}a_{\text{P}}\dfrac{\Gamma}{L_{\text{P}}}
 \label{eq:Type_I_Migration_rate}
\end{equation}
where $L_{\text{P}}=M_{\text{P}}\sqrt{GM_{\star}a_{\text{P}}}$ is the angular momentum of the planet, and the total torque is
\begin{equation}
 \Gamma = \left(1.364 + 0.541\dfrac{\text{dlog}\Sigma_{\text{g}}}{\text{dlog}a_{\text{P}}}\right)\left(\dfrac{M_{\text{P}}}{M_{\star}}\dfrac{a_{\text{P}}\Omega_{\text{k,P}}}{c_{\text{s,P}}}\right)^{2}\Sigma_{\text{g}}(a_{\text{P}})a^{4}_{\text{P}}\Omega^{2}_{\text{k,P}}.
 \label{eq:Type_I_Migration_rate2}
\end{equation}
Remember that the sub-index $\text{P}$ represents the location of the planet.

When a planet is massive enough to open a clear gap in the gas surface density profile, it undergoes type II migration. 
\citet{Crida2006} showed that a planet of mass $M_{\text{P}}$ is able to clean more than 90$\%$ of the gas around its orbit if
\begin{equation}
 \dfrac{3}{4}\dfrac{H_{\text{g}}}{R_{\text{H}}} + \dfrac{50\nu}{a^{2}_{\text{P}}\Omega_{\text{k,P}}}\left(\dfrac{M_{\star}}{M_{\text{P}}}\right) \leq 1.
 \label{eq:GapCriteria}
\end{equation}

Once the gap is opened, the orbital evolution of the planet is completely tied to the viscous evolution of the gas disc
and, as long as the local mass of the disc is greater than the mass of the planet \citep[``disc-dominated type II migration'' 
following][]{Armitage2007}, the planet migrates inward at a rate given by
\begin{equation}
  \left(\dfrac{\text{d}a_{\text{P}}}{\text{d}t}\right)_{\text{DD-migII}} = -\dfrac{3\nu}{2a_{\text{P}}},
 \label{eq:migII-rate-gasdisc}
\end{equation}
which is the same rate at which the gas moves in the disc \citep{IdaLin2004a}. 
When the mass of the planet is high enough to be comparable to the local mass of the disc \citep[``planet-dominated type II migration'' following][]{Armitage2007}, the migration rate is even lower and, as in \citet{Edgar2007} and \citet{Mordasini2009}, we calculate this
migration rate as
\begin{equation}
  \left(\dfrac{\text{d}a_{\text{P}}}{\text{d}t}\right)_{\text{PD-migII}} = -\dfrac{3\nu}{a_{\text{P}}}\dfrac{\Sigma_{\text{g}}(a_{\text{P}})a^{2}_{\text{P}}}{M_{\text{P}}}.
 \label{eq:migII-rate-gasdisc2}
\end{equation}
In this case, the migration rate not only depends on the viscosity of the disc but also on the gas surface density and on the 
mass of the planet.

\subsection{Water distribution on embryos and planetesimals}
\label{subsec:subsec11}

We also incorporate a water radial distribution for the embryos and planetesimals in our model of planet formation. In order to be consistent with the initial surface density of planetesimals adopted previously (Eq.~\ref{eq:solids} and \ref{eq:eta}), the initial fraction per unit mass of water in embryos and planetesimals is given by
\begin{equation}
  f_{\text{H}_2\text{O}}(R)= 
  \begin{cases}
    0,~\text{if}~R \le R_{\text{ice}}, \\
    \\
    \dfrac{\beta - 1}{\beta},~\text{if}~R > R_{\text{ice}}.
  \end{cases}
  \label{eq:water-distribution}
\end{equation}
The time evolution of the water radial distribution of the planetesimals is calculated as the Eq.~\ref{eq:densi-planetes-1} is solved. Moreover, the amount of water in the embryos is calculated self-consistently as they accrete planetesimals, and when two embryos merge between them we consider that the amount of water of the new embryo is the sum of the amounts of water of the previous ones. We remark that we do not consider loss of water either by the accretion of planetesimals or the merger between embryos. Thus, the final amounts of water represent upper limits.


\section{Initial conditions for our planetary systems}
\label{sec:sec12}

\begin{figure}
	\includegraphics[width=\columnwidth]{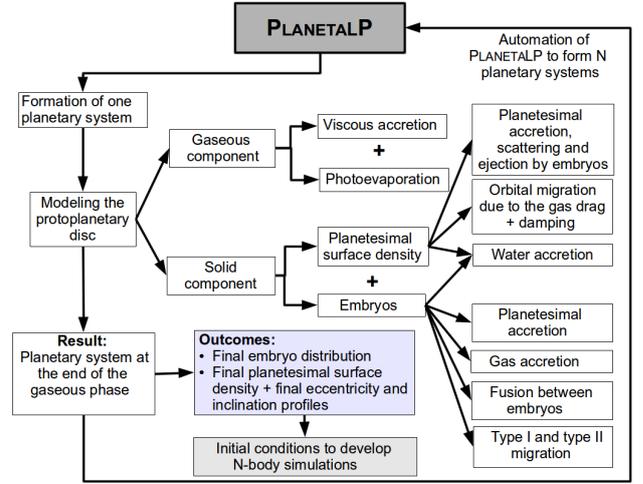}
    \caption{Schematic view of the operation of P{\scriptsize LANETA}LP.}
    \label{fig:2}
\end{figure}

The main goal of this work is to find favorable scenarios and disc properties for planetary systems similar to our own with 
P{\scriptsize LANETA}LP, to obtain initial conditions for the post-oligarchic growth of these systems (Ronco $\&$ de El\'{\i}a in prep., from now on, PII). 
To do so, P{\scriptsize LANETA}LP has been automated to develop several 
thousand planetary systems and to make a statistical study of population synthesis. Figure~\ref{fig:2} shows an schematic view of the 
operation of P{\scriptsize LANETA}LP.
In this section, we describe the initial conditions adopted to run P{\scriptsize LANETA}LP taking into account that we are not interested
in reproducing the actual exoplanet population, but we are interested in generating a great diversity of planetary systems without 
following any observable distribution in the disc parameters. 

\subsection{Scenarios and free parameters}
\label{subsec:subsec13}

As we have already mentioned, type I migration for idealized isothermal discs can be very fast, and thus, the timescales of migration are usually shorter than the disc lifetime. As in \citet{IdaLin2004a} and \citet{Miguel2011}, we then consider different scenarios for the type I migration introducing the reduction factor $f_{\text{migI}}$, which considers possible mechanisms that can slow down the planet migration 10 and 100 times ($f_{\text{migI}} = 0.1$ and $f_{\text{migI}} = 0.01$). We also consider the extreme cases in where type I migration is not slowed down at all ($f_{\text{migI}} = 1$), and where it is not considered ($f_{\text{migI}} = 0$). It is important to remark that we are not considering dynamical torques in our model \citep{Paardekooper2014}. These torques could slowed down significantly the migration rates estimated from the statical torques given by \citet{Tanaka2002}. \citet{Sasaki2017} found that the M-a diagram calculated from a simple planet population synthesis model, based in the work of \citet{IdaLin2004a} including statical and dynamical torques, can reproduce the observational exoplanet distribution around sun-like stars. It is important to note that this M-a diagram is similar to the M-a diagram obtained using statical torques and reduction factors in the type I migration rates.

Another important parameter for our model of planet formation is the size of the planetesimals. In the standard model of core accretion, both terrestrial planets and solid cores of gas giants are formed through the accretion of planetesimals \citep{Safronov1969, Wetherill1980, Hayashi1981}. However, the mechanisms for the formation of planetesimals and their initial sizes are still under debate \citep{Johansen2014, Testi2014, Birnstiel2016}. While collisional growth models could lead to the formation of sub-kilometer/kilometer planetesimals \citep{Okuzumi2012, Garaud2013, Kataoka2013, Drazkowska2013, Kobayashi2016, ArakawaNakamoto2016}, gravitational collapse models predict the formation of tens/hundreds kilometer planetesimals \citep{Johansen2007, Johansen2012, Cuzzi2008, Simon2016, Schafer2017}. Moreover, the size distribution of the asteroid belt can be reproduced by collisional evolution models using both, an initial population of big planetesimals \citep{Morbidelli2009} or an initial population of small planetesimals \citep{Weidenschilling2011}.   

In this work, as in \citet{Fortier2013}, we consider scenarios with different sizes for the planetesimals ($r_{\text{p}} = 100$~m, $1$~km, $10$~km and $100$~km), but we consider a single size per simulation. It is important to note that, while big planetesimals ($r_{\text{p}} \gtrsim 10$~km) are mainly governed by the quadratic regime for the determination of the radial drift and solid accretion rates, small planetesimals 
($r_{\text{p}} \lesssim 1$~km) can be in different drag regimes along the disc.

We consider random values from uniform distributions for all the free parameters because we are interested in developing a great diversity of planetary systems without following any observable parameter distribution. However, we take into account that the bounds of the parameter ranges are obtained from previous works. Following \citet{Andrews2009,Andrews2010}, we consider the mass of
the disc, $M_{\text{d}}$, between $0.01\text{M}_\odot$ and $0.15\text{M}_\odot$, the characteristic radius $R_{\text{c}}$ between 20~au and 50~au, and
the density profile gradient $\gamma$ between 0.5 and 1.5. The factor $\beta$ is taken randomly between 1.1 and 3 given that,
in our solar system, \citet{Lodders2003} usually takes a value of approximately 2 and \citet{Hayashi1981} adopts a value of 4. 
It is important to note that a value of $\beta = 1.1$ represents an amount of 9\% of water by mass, 
while a value of $\beta = 3$ represents an amount of 66\% of water by mass on embryos
and planetesimals at the beginning of the simulations and beyond the snowline, following Eq. \ref{eq:water-distribution}. 
The viscosity parameter $\alpha$ for the disc and the rate of EUV ionizing photons $f_{41}$
have a uniform distribution in log between $10^{-4}$ and $10^{-2}$, and between $10^{-1}$ and $10^{4}$, respectively \citep{DangeloMarzari2012}.

Since the maximum value of the mass of the discs considered is relatively high and since this can lead to the occurrence 
of gravitational instabilities, our model checks, according to the Toomre criterion \citep{Toomre1964} given by
\begin{equation}
  Q = \dfrac{c_{\text{s}}\Omega_{\text{k}}}{\pi G \Sigma_{\text{g}}},
 \label{eq:migII-rate-gasdisc2}
\end{equation}
the stability of our discs. When $Q > 1$ discs are considered unstable. It is important to remark that all the considered
discs in our simulations result to be stable throughout their entire extent.

Other parameters are kept constant for all the simulations. As we are particularly interested in the formation of solar 
system analogs, the mass of the central star is always $1\text{M}_\odot$ and the metallicity is $[\text{Fe}/\text{H}]=0$. Also the 
planetesimal and embryo densities are considered constant and to be 1.5 g~cm$^{-3}$ and 3 g~cm$^{-3}$, respectively. 

Regarding the embryo population, we consider an embryo distribution between 0.1~au and 30~au. 
These embryos are separated by 10 mutual Hill radii assuming circular and coplanar orbits, and their initial masses, 
which are of the order of the mass of the Moon, are given by the transition mass between the runaway and the oligarchic regime 
\citep{IdaMakino1993} as
\begin{equation}
 M_{\text{oli}} = 1.6R^{6/5}10^{3/5}m^{3/5}_{\text{p}}\Sigma^{3/5}_{\text{p}}M^{-1/5}_\star
  \label{eq:transitionmass}
\end{equation}
where $m_{\text{p}}$ is the mass of the planetesimals.

Finally, with the aim of finding the most suitable scenarios and parameters that provide us with solar system analogs we perform
1000 simulations per each combination between $f_{\text{migI}}$ and $r_{\text{p}}$. This is, we form 16000 different planetary systems
that evolve in time until the gas of the disc dissipates. From now on, when we talk about formation scenarios, 
we are referring to all the combinations between $f_{\text{migI}}$ and $r_{\text{p}}$.

\subsection{The disc dissipation timescale}
\label{subsec:subsec14}

Another important parameter that determines the final configuration of a 
planetary system during the gaseous phase is the 
gas-disc dissipation timescale, $\tau$. \citet{Haisch2001} and \citet{Mamajek2009} have observed young stars in
clusters of different ages and they suggested that protoplanetary discs have characteristic lifetimes between 3~Myr and 10~Myr. 
More recently, \citet{Pfalzner2014} showed that, taking into account that the previous works present observational biases, 
gas discs could last much longer than 10~Myr.

In our model, unlike \citet{Miguel2011} for example, wherein the evolution of the gas is represented by an exponential decay law, 
the gas disc evolves due to the viscous accretion and photoevaporation processes.
Both phenomena together determine the lifetime of the gas, $\tau$. Therefore, we need to form planetary systems whose combinations
between the viscosity parameter $\alpha$, the EUV rate $f_{41}$, the mass of the disc $M_{\text{d}}$, the characteristic 
radius of the disc $R_{\text{c}}$, and the surface density gradient $\gamma$ make the gas discs of these systems to
dissipate on timescales according to the results of the previous mentioned works. To do this, we run a version of 
P{\scriptsize LANETA}LP that is limited to the study of the evolution in time of the gas disc. This is, we ``turn off''
the evolution of the planetesimal surface density and the evolution of the embryo population.
We then generate, using a von Neumann acceptance-rejection method, as many gas discs with dissipation timescales between 1 and 12~Myr as we need, this is, 16000 combinations of all the parameters to develop complete planetary systems.

\begin{figure}
	\includegraphics[width=\columnwidth]{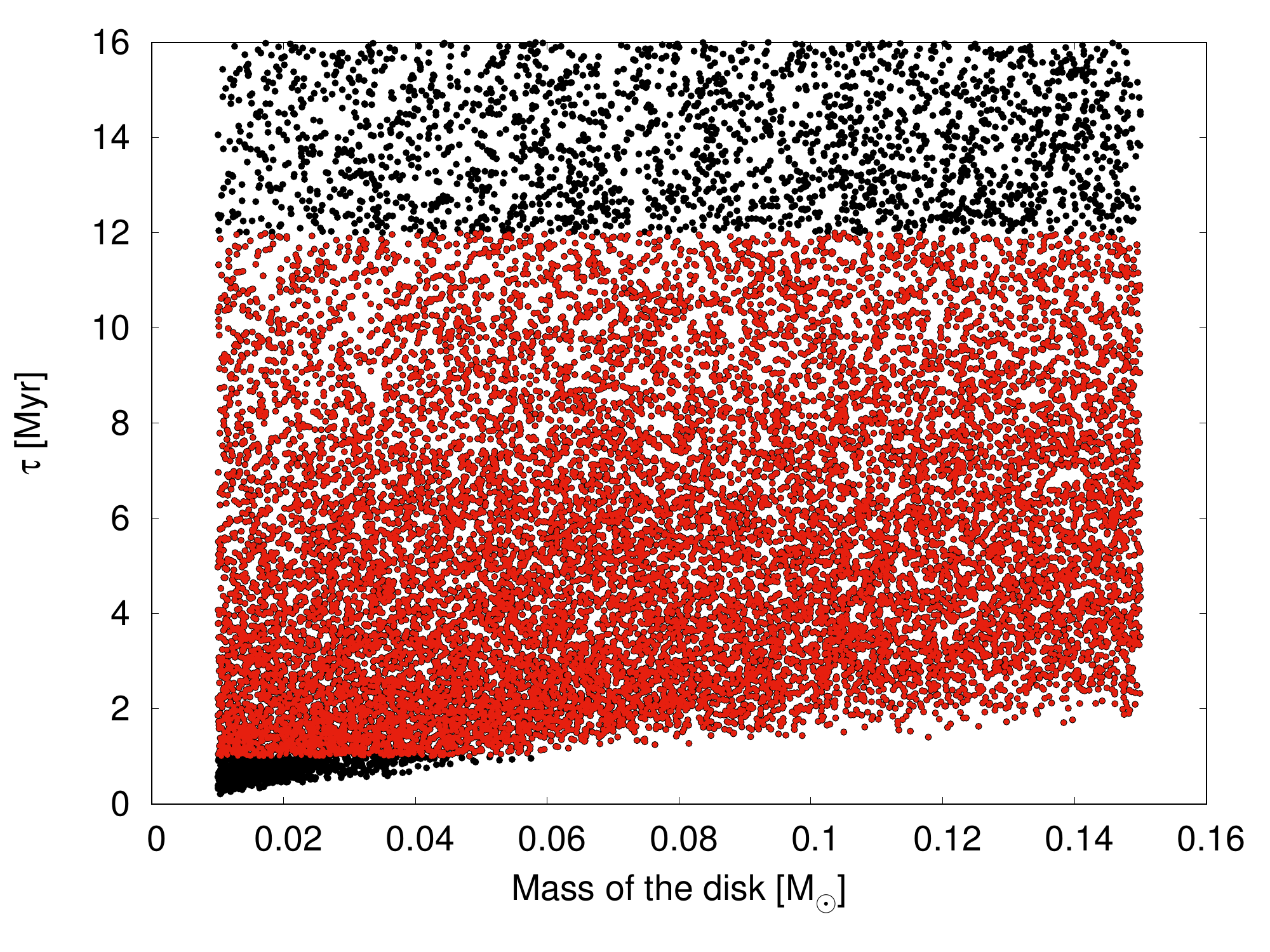}
    \caption{Dissipation timescale in~Myr as a function of the mass of the disc in solar masses for all the simulations performed. 
      Each point represents a set of parameters randomly taken that is going to build a final planetary system. The red points represent 
      those sets whose combinations between all the parameters managed to obtain a dissipation timescale $\tau$ between 1 and 12~Myr. 
      The black points are those sets that did not managed to obtain a suitable $\tau$ and are not going to be taken into account.}
    \label{fig:3}
\end{figure}

Fig \ref{fig:3} shows the timescale (in~Myr) vs. mass of the disc (in solar masses) plane. Each point represents a set of parameters 
randomly taken. The 16000 red points represent sets whose combinations between all the parameters obtained a disc of gas that dissipated 
between 1~Myr to 12~Myr. We find a mean value of $\tau$ of 6.45~Myr with a dispersion of $\sigma= 2.67$~Myr. The black points are those sets that resulted in discs dissipating in less than 1~Myr or in more than 12~Myr. These last sets are then discarded for our simulations.

\section{Results}
\label{sec:sec15}

The main goal of this work is to find, through the population synthesis analysis developed with our model of planet formation, 
sets of appropriate initial conditions to study, in a future work (PII), the post-oligarchic evolution 
of planetary systems similar to our own with N-body simulations. However, through the development of the 16000 simulations, 
which were performed randomly and uniformly varying the parameters of the disc, 
we found a great diversity of planetary systems architectures, not only solar system analogs.

\subsection{General results}

To classify the diversity of planetary systems, 
we define general properties
for the different types of planets. On the one hand, gas giant planets present an envelope which is greater 
than the mass of the core. Within this category we can find Jupiter-like or Saturn-like planets. The difference between them is that 
we consider that Jupiter-like planets manage to open a gap in the disc or their total mass is greater than $200\text{M}_\oplus$. 
On the other hand, icy giant planets present envelopes greater 
than 1\% of the core mass but must have envelopes smaller than the core mass. At last, rocky
planets present envelopes smaller than 1\% of their core mass. Having classified the final planets, from all the 
diversity of planetary systems developed, we can distinguish three global kinds of planetary systems that we classify as:
\begin{itemize}
\item {\bf Rocky planetary systems}: planetary systems which are formed only by rocky planets, without gas or icy giants. 
\item {\bf Icy giant planetary systems}: these planetary systems can harbor rocky planets but they must present at least one icy giant planet
analog to a Neptune-like planet, or a hot/warm Neptune-like planet.
\item {\bf Gas giants planetary systems}: planetary systems which must harbor at least one giant planet (in this global characterization 
we do not distinguish the location of the giant), but can also contain rocky and icy giant planets.
\end{itemize}

In this work we are particularly interested in those planetary systems which, at the end of the gas phase, present similarities to our 
own solar system, and so we classify them as:
\begin{itemize}
\item {\bf Solar system analogs (SSA)}: planetary systems that host only rocky planets in the inner zone of the disc and at least one 
giant planet beyond 1.5~au.
\end{itemize} 
SSA represent a subkind of those planetary systems with gas giants. Table \ref{tab:1} shows the general percentages of each kind, 
including the percentage that SSA represent respect to the total of the performed simulations. Although the scope of this work is 
to focus only on SSA, we will mention some general results of the developed simulations that are in good agreement with 
previous works.

\begin{table}
	\centering
	\caption{Final percentages of planetary systems with respect to the total of the developed simulations.}
	\label{tab:1}
	\begin{tabular}{lc} 
		\hline
                Planetary system & Final percentage \\
                \hline
                \hline
		{\bf Rocky planetary systems} & 64.46\% \\
                {\bf Icy giant planetary systems} & 13.4\% \\
                {\bf Gas giant planetary systems} &  22.14\% \\
		\hline
                {\bf Solar system analogs} & 4.3\% \\
                \hline
                \hline
	\end{tabular}

\end{table}

Following Table \ref{tab:1} it is clear that rocky planetary systems represent the vast majority, indeed, more than 60\% of the total of the
simulated planetary systems. 
Moreover, we can find this kind of systems in all the formation scenarios considered. In fact, rocky planetary systems represent
more than 40\% of the planetary systems in each formation scenario for all of them but one, wherein represent $\sim 20\%$. Finally, we can 
also find them in the whole range of disc masses considered, between $0.01\text{M}_\odot$ and $0.15\text{M}_\odot$. This result is in good agreement with 
the results obtained by \citet{Miguel2011} although they followed observable distributions for the most important disc parameters and 
although they only considered planetesimals of 10~km.

\begin{figure*}
	\includegraphics[angle=270, width=0.9\textwidth]{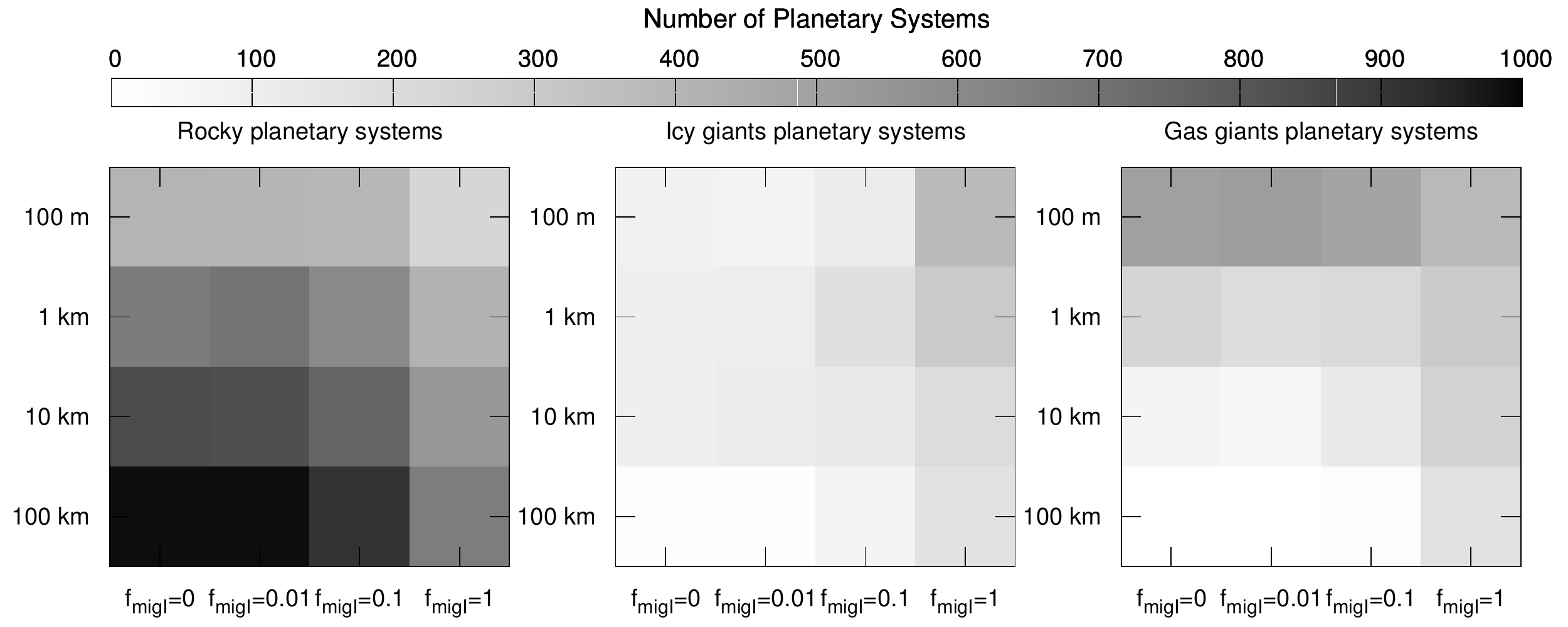}
    \caption{Maps of density of rocky planetary systems (left), icy giant planetary systems (middle) and gas giant planetary systems (right). 
The greyscale represents the number of planetary systems in each formation scenario. For each scenario, the sum of rocky, icy giant and gas giant planetary systems is equal to 1000. }
    \label{fig:4}
\end{figure*}

Fig.~\ref{fig:4} shows maps of density of the different planetary systems for each formation scenario.
Rocky planetary systems are 
significantly more favorable in formation scenarios with big planetesimals and low-moderate/null type I migration rates, although,
as we mentioned before, we find them in each of all the formation scenarios. This is a natural 
consequence of the fact that planetesimal accretion rates are smaller for big planetesimals, and thus, the formation of massive 
cores generally requires formation timescales that exceed the characteristic dissipation timescales of the discs. Besides, the smaller 
the type I migration rate, the smaller the probability of merger between embryos and thus, the formation of massive cores by this 
mechanism. It is also important to note that a significant number of rocky planetary systems is formed by 
small planetesimals and low-moderate/null type I migration rates. This is due to the fact that Fig.~\ref{fig:4} does not represents 
explicitly the dependence with the mass of the discs. As we adopted that the mass of the disc is uniformly distributed between $0.01\text{M}_\odot$ and $0.15\text{M}_\odot$,
there is a significant percentage of systems in which the total mass of the disc is not enough to form massive cores, and consequently, icy giant or gas giant planets.

The opposite case is given by the icy giant planetary systems. These planetary systems are mainly formed in scenarios with small 
planetesimals and large type I migration rates. The planetesimal accretion rates are greater for smaller planetesimals, situation that 
lead to the quickly formation of several Earth-cores beyond the snowline. However, large type I migration rates produces that these 
cores quickly reach the inner edge of the disc, avoiding the formation of gas giants. Therefore, on the one hand, high type I migration rates
could lead to the merge of embryos, an thus, the formation of massive cores as we mentioned before. But, on the other hand, high type I 
migration rates also cause the planets to quickly get into the inner zone, and fail in the gas giant formation.

Finally, gas giant planetary systems are majority for small planetesimals and low-moderate/null type I migration rates. As it was already mentioned, small planetesimals favor the formation of massive cores, and low-moderate/null type I migration rates avoid the planet to drop into the inner edge of the disc, allowing the planet to trigger the gaseous runaway growth. However it is important to note that we find gas giants in scenarios with high type I migration rates. About a $70\%$ of the planetary systems
  formed in scenarios with high type I migration rates, considering all the chosen planetesimal sizes, present Hot/Warm-Jupiters within 0.5 au. However, taking into account all the formation scenarios that form giant planets, not only those with high type I migration rates, the planetary systems with Hot/Warm-Jupiters represent the 43\%
  of Gas Giant planetary systems. Moreover, those planetary systems with Hot/Warm-Jupiters represent only a 9.5\% of the total of the developed simulations. Thus,
  although we find higher percentages of Hot-Jupiter occurences than those predicted observationally, which are $\sim 1\%$ \citep{Marcy2005,Cumming2008,Mayor2011,Wright2012}, \citet{Wang2015} concluded that the current knowledge of stellar properties and the stellar multiplicity rate is yet very limited to reach
quantitative results for the Hot-Jupiter occurrence.

We remark again, that these density maps do not take into account the dependency with the properties of the disc, i.e. the mass of the discs, the exponent $\gamma$, the characteristic radius, and the factor $\beta$. 

\subsection{Solar system analogs}
\label{subsec:subsec16}

\begin{figure*}
	\includegraphics[angle=270, width=0.8\textwidth]{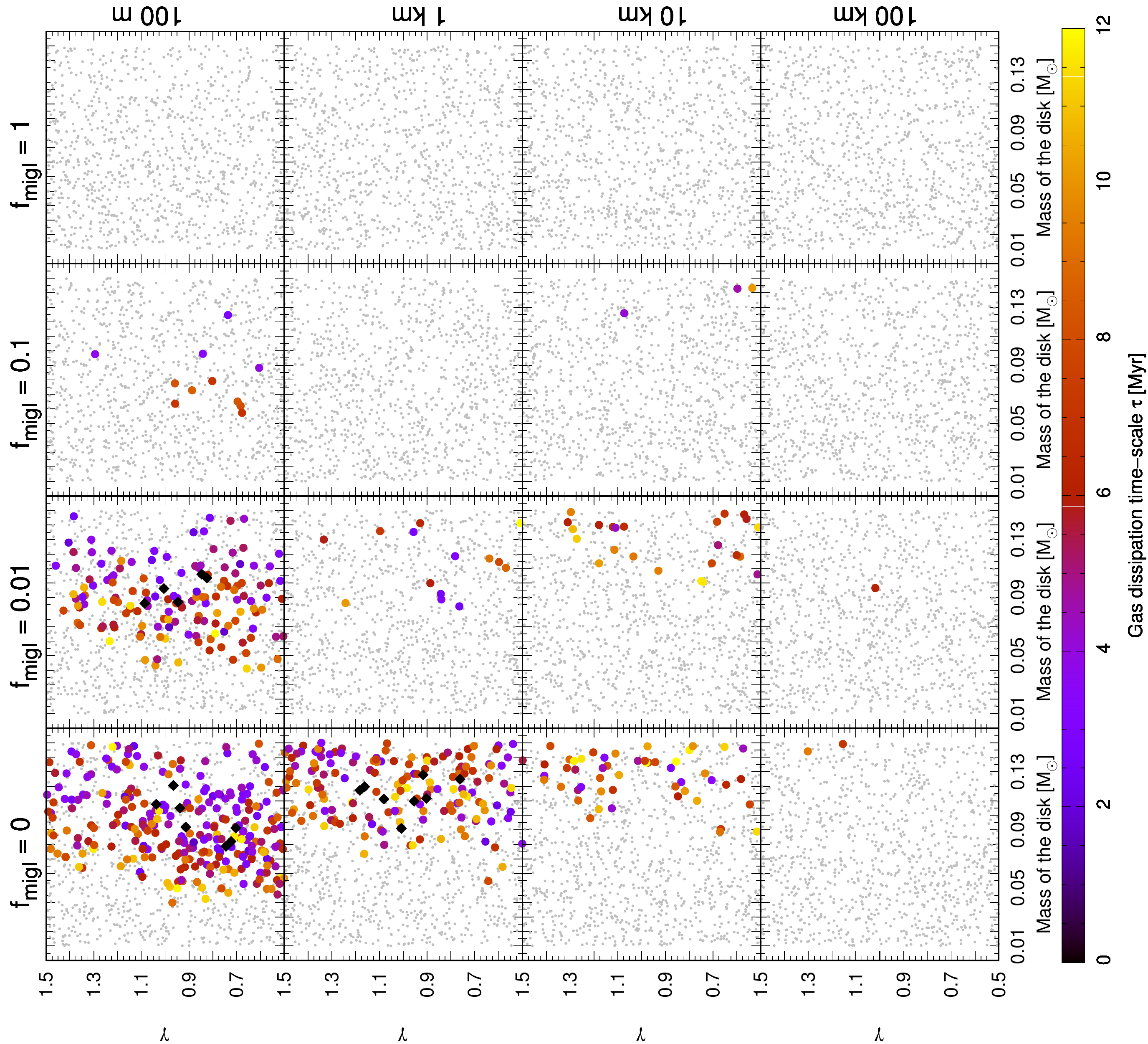}
        \caption{Distribution of the planetary systems produced by P{\scriptsize LANETA}LP in a plane $M_{\text{d}}$ - $\gamma$.
          SSA are represented with colored points. The colorscale represents the dissipation timescale of each disc. Grey points represent
          the rest of the planetary systems that are not SSA. Black diamonds are those SSA with all their disc parameters between $\pm \sigma$.
          Each row and each column shows the results for the different formation scenarios.}
    \label{fig:6}
\end{figure*}

Of the performed simulations, we found that only 688 protoplanetary discs formed SSA, this represents only a 4,3\% of the total. 
Table \ref{tab:2} shows the percentages of SSA for each formation scenario and Figure~\ref{fig:6} shows $\gamma$ vs. $M_{\text{d}}$ planes for 
each formation scenario. In this figure, each point represents a formed planetary system with a particular set of disc parameters. 
The colored points represent SSA, the grey points are the rest of the planetary systems performed in which we are not going to focus in 
this work. The colorscale shows the dissipation timescale of the gas disc for each planetary system. 
As we can see in table \ref{tab:2} and figure \ref{fig:6}, we do not find SSA in scenarios with non-reduced type I migration. 
This is due to the fact that, as we have already mentioned, type I migration without any 
reduction factor causes that giant-forming planets in the outermost zone of the disc migrate to within 1.5~au in timescales smaller than the 
dissipation of the gas disc, regardless of the size of the planetesimals considered. Clearly, formation scenarios of low (or none) 
type I migration rates and small planetesimals (100~m and 1~km) are the most favorable for the formation of SSA. Since the planetesimal
accretion rates are higher for small planetesimals, the rapid 
formation of massive cores before the dissipation of the gas is favored. Otherwise, for large planetesimals, the formation timescales
of massive nuclei are not high enough to trigger the growth of the gaseous envelope and to form giant planets in timescales according to
the disc lifetime. 

\begin{table*}
	\centering
	\caption{Percentage of SSA found in each formation scenario.}
	\label{tab:2}
	\begin{tabular}{ccccc} 
		\hline
                Planetesimal size & $f_{\text{migI}}=0$ & $f_{\text{migI}}=0.01$ & $f_{\text{migI}}=0.1$ & $f_{\text{migI}}=1$ \\
                \hline
                \hline
		100~m  & 26.90\% & 14.90\% & 1.10\% & 0\% \\
		1~km   & 16.40\% & 1.4\% & 0\% & 0\% \\
		10~km  & 5.10\% & 2.40\% & 0.3\% & 0\% \\
		100~km & 0.2\% & 0.1\% & 0\% & 0\% \\
		\hline
	\end{tabular}
\end{table*}

\subsubsection{Solar system analogs architectures}
\label{subsubsec:subsubsec17}

\begin{figure*}
	\includegraphics[angle=270, width=0.8\textwidth]{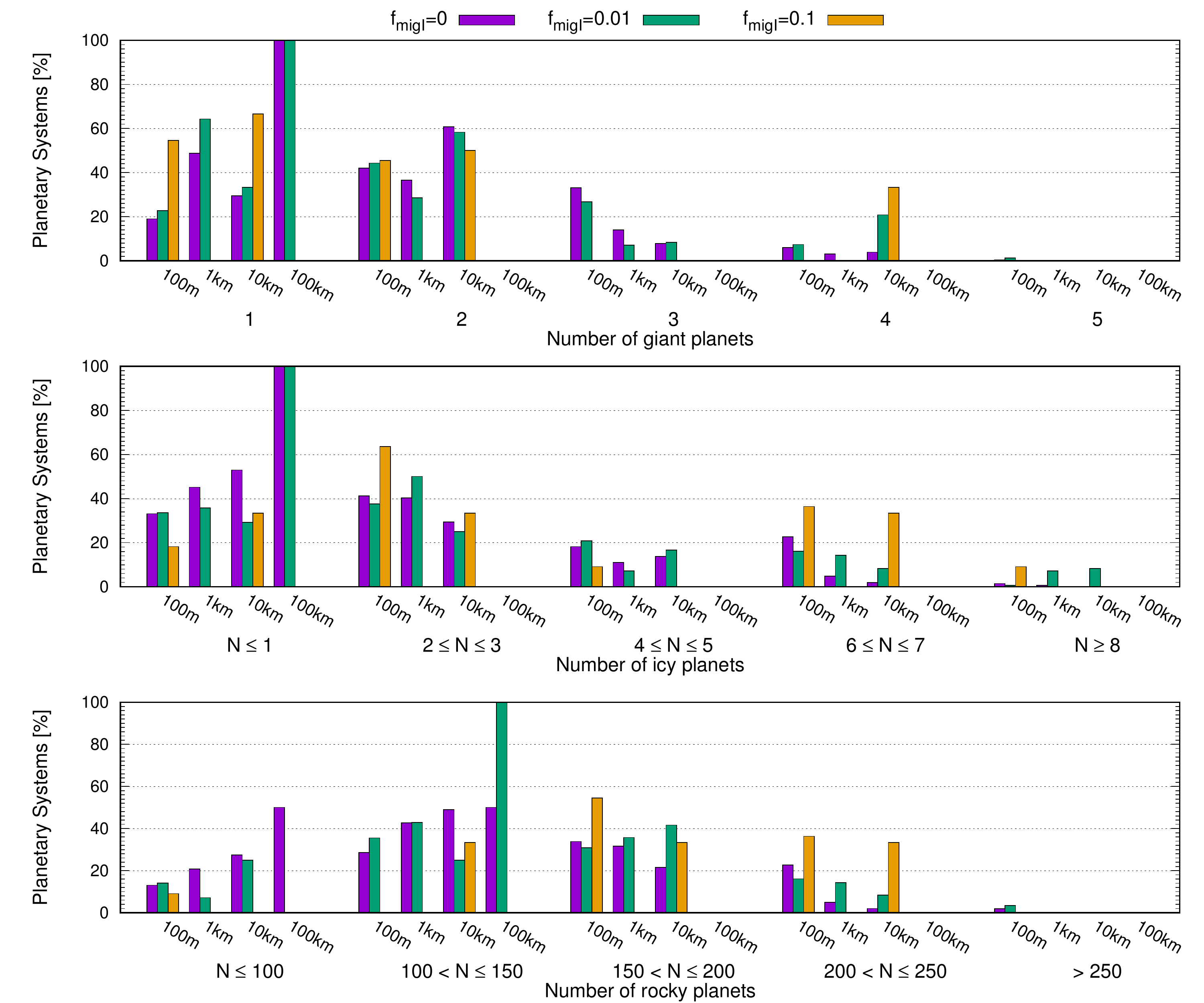}
    \caption{Histograms showing how many gas giants, icy giants and rocky planets present each SSA formation scenario.}
    \label{fig:5}
\end{figure*}

As we already mentioned, SSA must harbor at least one giant planet beyond 1.5~au and rocky planets in the inner regions of the disc. However, the architectures of the SSA found are quite
different, and not all of them present icy giants or only one gas giant planet. Taking into account all the SSA, without distinguishing between planetesimal sizes or
migration rates, the most representative architectures found among the SSA are those that present between 1 to 3 gas giants, between 0 and 4 icy giants and between 100 and 200 rocky planets
along the disc. In fact, the mean of the number of gas giants is 2, with a dispersion of $\sim 1$, the mean of the number of icy giants is also 2, with a dispersion of $\sim 2$,
and the mean of the number of rocky planets is 150, with a dispersion of $\sim 47$.
Particularly, Figure~\ref{fig:5} displays histograms showing the percentage of SSA that present one, two, three, four or five gas giants, and also shows how many
systems present different numbers of icy giants and rocky planets, but differentiating the systems according to the size of planetesimals and migration rates. 
For planetesimals of 100~km, only one giant planet is formed in the only three systems found for low or null type I migration. These systems also harbor only one, or
none, icy giant, but a significant number of rocky embryos, some of them in the outer part of the disc. For these big planetesimals, we do not find SSA formed
with moderate and non reduced type I migration rates.
The formation of two giant planets is more likely for low or null type I migrations rates and scenarios with planetesimals of 10~km, while the formation of only one gas giant is more favorable for moderate type I migration rates. It is important to note that, in both cases, there is a not negligible probability to form systems with three or four giant planets.
These systems also harbor preferentially a small number of icy giants. However, we found that for the case of null or low type I migration rates, it is possible to find SSA that
harbor until 8 icy giants. These kind of systems also present preferably between 100 and 200 rocky embryos. For small planetesimals of 1~km and 100~m we find similar results.
In both cases, planetary systems prefentially harbor one or two giant planets, with the difference that no giant planet is formed for planetesimals of 1~km of radius for moderate
type I migration rates. In both cases, a small number of icy giants is also expected. Finally, a large number of rocky planets are expected in such systems when the gas of the disc is dissipated. It is important to remark that the stability and dynamical 
evolution of these kind of systems after the gas dissipation will be discussed in the forthcoming paper.

\subsubsection{Favorable disc parameters}
\label{subsubsec:subsubsec18}

Taking into account all the SSA, table \ref{tab:3} shows the ranges, mean and dispersion values for the disc parameters. As we can see, we can find SSA for all the ranges
considered of $\gamma$, $R_{\text{c}}$, $\alpha$ and $\tau$. However, we do not find SSA for low-mass discs ($M_{\text{d}} < 0.04\text{M}_\odot$) because low-mass discs have not enough solid material to form massive cores before the disc dissipation timescales. Similar results were found by \citet{Thommes2008} and \citet{Miguel2011}. Although we find
SSA for almost all the values in the ranges of the free disc parameters, they are preferentially formed for massive disc ($<M_{\text{d}}>= 0.1~\text{M}_\odot$), smooth surface densities
profiles ($<\gamma>= 0.95$), moderate values for the disc characteristic radius ($<R_{\text{c}}> \sim 33$~au), small values of the $\alpha$-viscosity parameter
($<\alpha>= 3.4 \times 10^{-4}$), and moderate timescales for the disc dissipation ($<\tau> \sim 6.5$~Myr).
It is worth mentioning that only 20 SSA have all their disc parameters between the dispersion values, $\pm \sigma$. These planetary systems are represented with black diamonds
in Figure~\ref{fig:6}. This represents only a $\sim 3~\%$ of the total SSA. All these systems are formed considering small planetesimals (100~m and 1~km) and adopting null or low type
I migration rates.

\begin{table}
	\centering
	\caption{Ranges, means and dispersion values of the disc parameters that formed SSA.}
	\label{tab:3}
	\begin{tabular}{cccc} 
		\hline
                Disc & Range & Mean & $\sigma$ \\
                parameter & & & \\
                \hline
                \hline
		$M_{\text{d}}$  & $0.04\text{M}_\odot$ - $1.15\text{M}_\odot$ & $0.10\text{M}_\odot$ & $0.027\text{M}_\odot$   \\
		$\gamma$   & $0.5$ - $1.5$ & 0.95 & 0.27  \\
		$R_{\text{c}}$  & 20~au - 50~au & 32.8~au & 8.73~au  \\
		$\alpha$ & $10^{-4}$ - $10^{-2}$ & $3.4\times10^{-3}$ & $2.8\times10^{-3}$  \\
		$\tau$ & 1.65~Myr - 11.96~Myr & 6.45~Myr & 2.67~Myr   \\
                \hline
	\end{tabular}
\end{table}

\subsubsection{Evolution and final configuration of solar system analogs}
\label{subsubsec:subsubsec19}

In this section we describe the general characteristics of the evolution in time of a SSA, and also show the final 
configurations of the most representative SSA of each formation scenario, which are composed of:
\begin{itemize}
\item An embryo distribution that provide us with information about the semimajor-axis, mass of the core, mass of the envelope, 
mass of water due to the accretion of planetesimals and mass of water due to the fusion with other embryos, of each final body.
\item A planetesimal surface density, eccentricity and inclination profiles.  
\end{itemize} 
These final configurations are going to serve as initial conditions to analyze the post-oligarchic stage of planet formation, 
via N-body simulations (PII).

\begin{figure*}
	\includegraphics[angle=270, width=\textwidth]{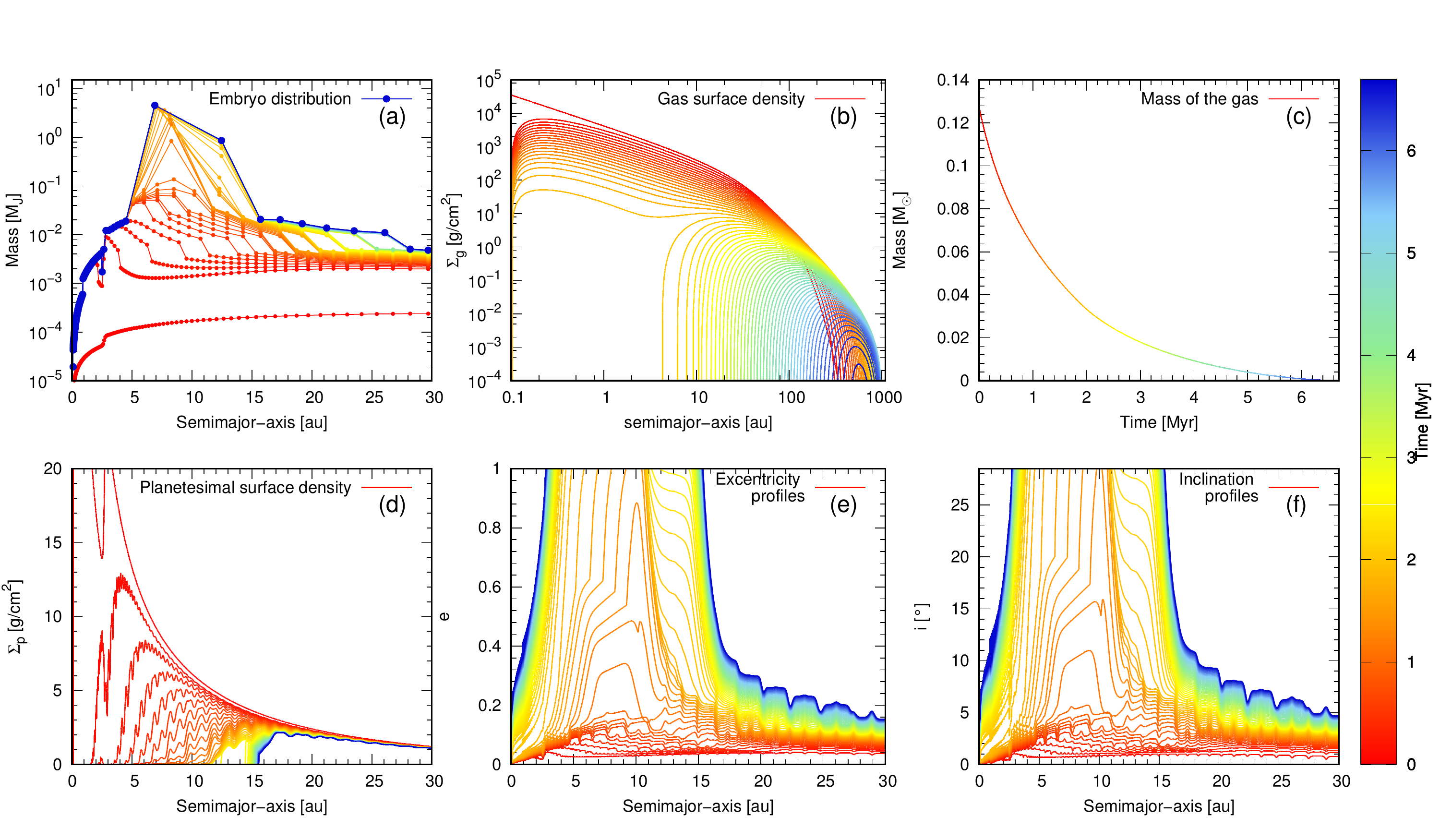}
    \caption{Example of the evolution in time of the embryo distribution (a), the gas surface density (b), the mass of the disc (c), 
      the planetesimal surface density (d) and the eccentricity and inclination profiles (e and f, respectively) of a SSA with 
planetesimals of 1~km and without type I migration. The parameters of this disc are: $M_{\text{d}}=0.13\text{M}_\odot$, $\gamma=0.92$, $R_{\text{c}}=34$~au, $\alpha=1.1\times10^{-3}$ and $\tau=6.65$~Myr. Finally $\beta=1.96$. The colorscale represents the time
      evolution of the system, and each profile is plotted every 0.1~Myr.}
    \label{fig:7}
\end{figure*}

Figure~\ref{fig:7} shows the evolution in time, represented by the colorscale, of a representative SSA for the formation scenario without 
type I migration and with planetesimals of 1~km. The parameters of this particular system are $M_{\text{d}}=0.13\text{M}_\odot$, $\gamma=0.92$, 
$R_{\text{c}}=34$~au, $\alpha=1.1 \times 10^{-3}$, and a dissipation timescale of $\tau=6.65$~Myr. The factor that represents the amount 
of solid material condensed beyond the snowline, at 2.7~au, is $\beta=1.96$, in this example. Panel (a) represents the
time evolution of the embryo distribution. The small red points linked by lines represent the initial embryo distribution.
These embryos start to grow due to the accretion of planetesimals within their feeding zones. In time, as they grow, if the distance
between them is less than 3.5~Hill radius they merge into one object. When the cores reach a critical mass, which is approximately 
$10\text{M}_\oplus$, they start to accrete significant amounts of gas until they achieve their final masses and the disc is dissipated. 
Panels (b) and (c) represent the evolution in
time of the gaseous component of the disc. The gas surface density profile (b), which is plotted every 0.1~Myr, begins to decrease
and to expand in the radial direction due to the disc angular momentum conservation. At approximately 2~Myr, a gap is opened in the 
disc due to the photoevaporation process. As a consequence, the gas in the inner zone of the disc, inside the gap, rapidly falls 
to the central star, while the gas in the outer zone continues its evolution for a few million years. The mass of the gas in the disc 
decreases as it is shown in panel (c). The planetesimal surface density (d) evolves in
time due to the radial drift and due to the embryos, which accrete and excite them, increasing their eccentricities (panel e) and inclinations (panel f). As time advances, the planetesimal profile decreases, and the inner zone of the disc quickly runs out of planetesimals. 
This is mainly due to the fact that the planetesimal
accretion rates are higher in the inner zone where the planetesimal surface density is initially higher. This favors the formation of 
many, low-mass rocky planets very close to each other, which accrete all the available solid material. These embryos do not merge between 
them due to the fact that their feeding zones are very narrow, an thus, they do not grow enough to be separated less than 3.5 mutual Hill 
radius (specially, very close to the star). Besides, the formation of two giant planets (see panel a)
helps to remove planetesimals between 10~au and 15~au. At the end, when there is no more gas in the disc, there are only planetesimals 
behind 15~au. This remnant of planetesimals (represented with the blue line) present final eccentricity and inclination profiles according to the blue lines in panels (e) and (f), respectively. It is important to note that these eccentricity and inclination profiles along the 
disc have physical sense only in the region where there is still presence of planetesimals, for this particular case, beyond $\sim$ 15~au.

\begin{figure*}
	\includegraphics[angle=270, width=\textwidth]{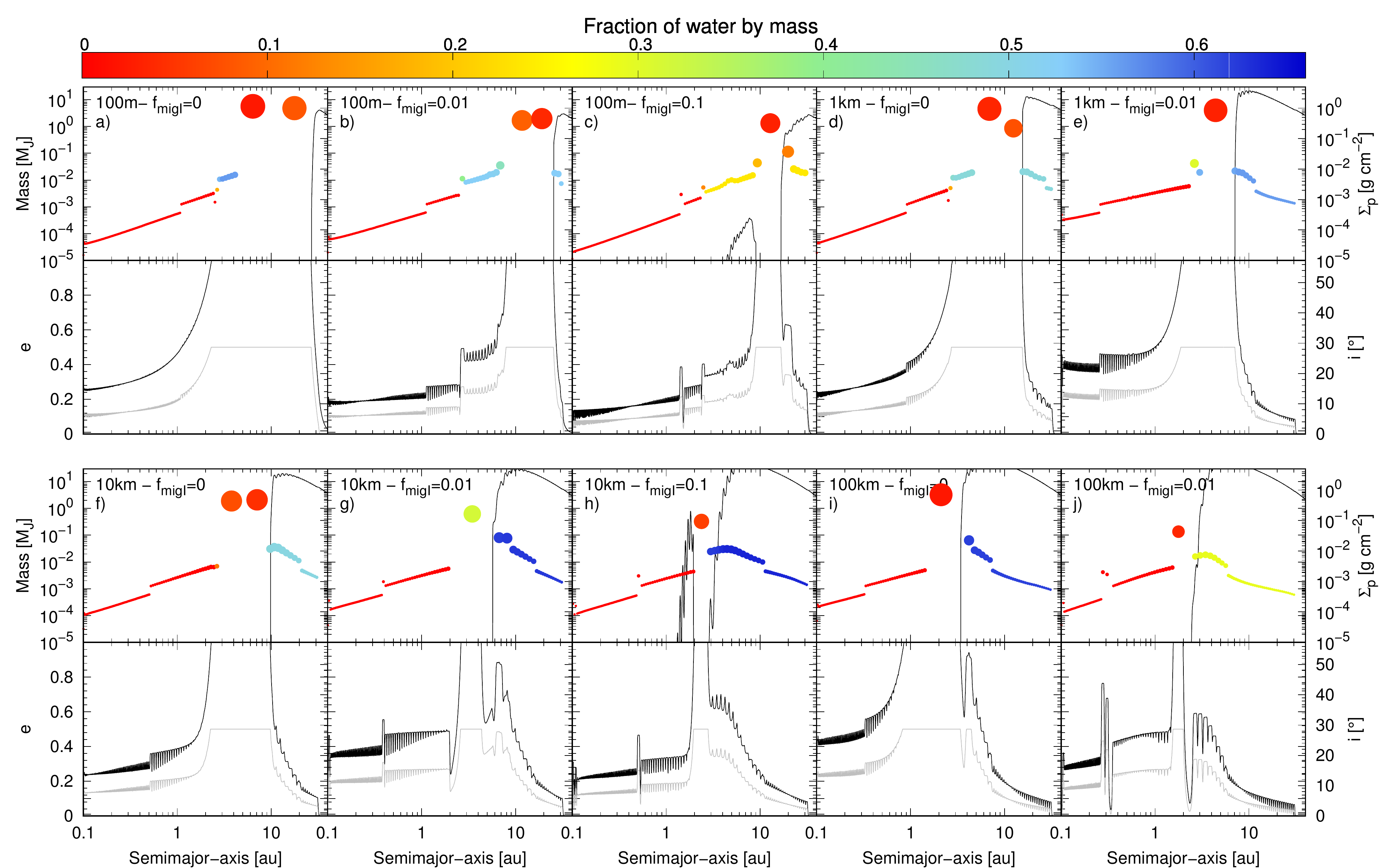}
    \caption{Representative planetary configurations at the end of the gaseous phase of all the formation scenarios that present SSA in a mass vs. semimajor-axis vs. 
      planetesimal surface density plane, and eccentricity vs. semimajor-axis vs. inclination profile plane.
      Each colored point represents a final planet, the solid black line in all
      the planetary systems represents the final planetesimal density profile at that time. Under these planet distributions and planetesimal profiles we can see the final
      planetesimal eccentricity and inclination profiles, in black and grey lines, respectively.
      These final profiles are only important for those regions in the disc where there is still a planetesimal population.
      The colorscale represents the final amounts of water by mass in each planet and the size of each planet is represented
      in logarithmic scale. The final planetary system obtained with planetesimals of 1~km and without type I migration (panel d) 
is the same planetary system exemplified on Figure~\ref{fig:7}.}
    \label{fig:8}
\end{figure*}

As we mentioned before, the aim of this work is to find suitable initial conditions for SSA to develop N-body simulations, thus,
Figure~\ref{fig:8} shows representative final configurations of SSA in all the formation scenarios in which they were found. 
Each panel shows the final 
embryo distribution with the final planetesimal density profile of each system at the top, and the final
planetesimal eccentricity and inclination profiles in the bottom. The colorscale represents the percentage of water by mass.

Although all this planetary systems are SSA, they present global differences regarding on the location of the giant planets and on the final
planetesimal surface density profiles. On the one hand, as the size of the planetesimals and the reduction factor of type I migration 
increase, the location of the giant planets changes, moving inward. On the other hand, also the planetesimal remnant until 30~au 
increases towards the central star.
The final location of the giant planet depends on several phenomena. As it was shown by \citet{Guilera2011}, the competition
between the planetesimal accretion timescales and the planetesimal migration timescales, regulates the formation of a giant planet. These 
authors found that these two timescales depend on the planetesimal size and on the location of the planet with respect to the central star. 
Thus, the optimal location in the disc for the formation of a massive core, precursor of a giant planet, is different for different 
planetesimal sizes. Besides, the main properties of the protoplanetary disc, i.e. the slopes of the surface densities, the jump in the 
position of the snowline due to the condensation of volatile material, the characteristic radius, etc., can play an important role in the
formation of a giant. In fact, \citet{Guilera2011} found that for smooth slopes, the formation of a giant planet is optimized in the outer
part of the disc for scenarios with small planetesimals ($r_{\text{p}} < 100$~m). Moreover, another important phenomenon that could change the
optimal location of the formation of a giant planet, is the merger between several planets.

\begin{figure}
	\includegraphics[angle=270, width=0.95\columnwidth]{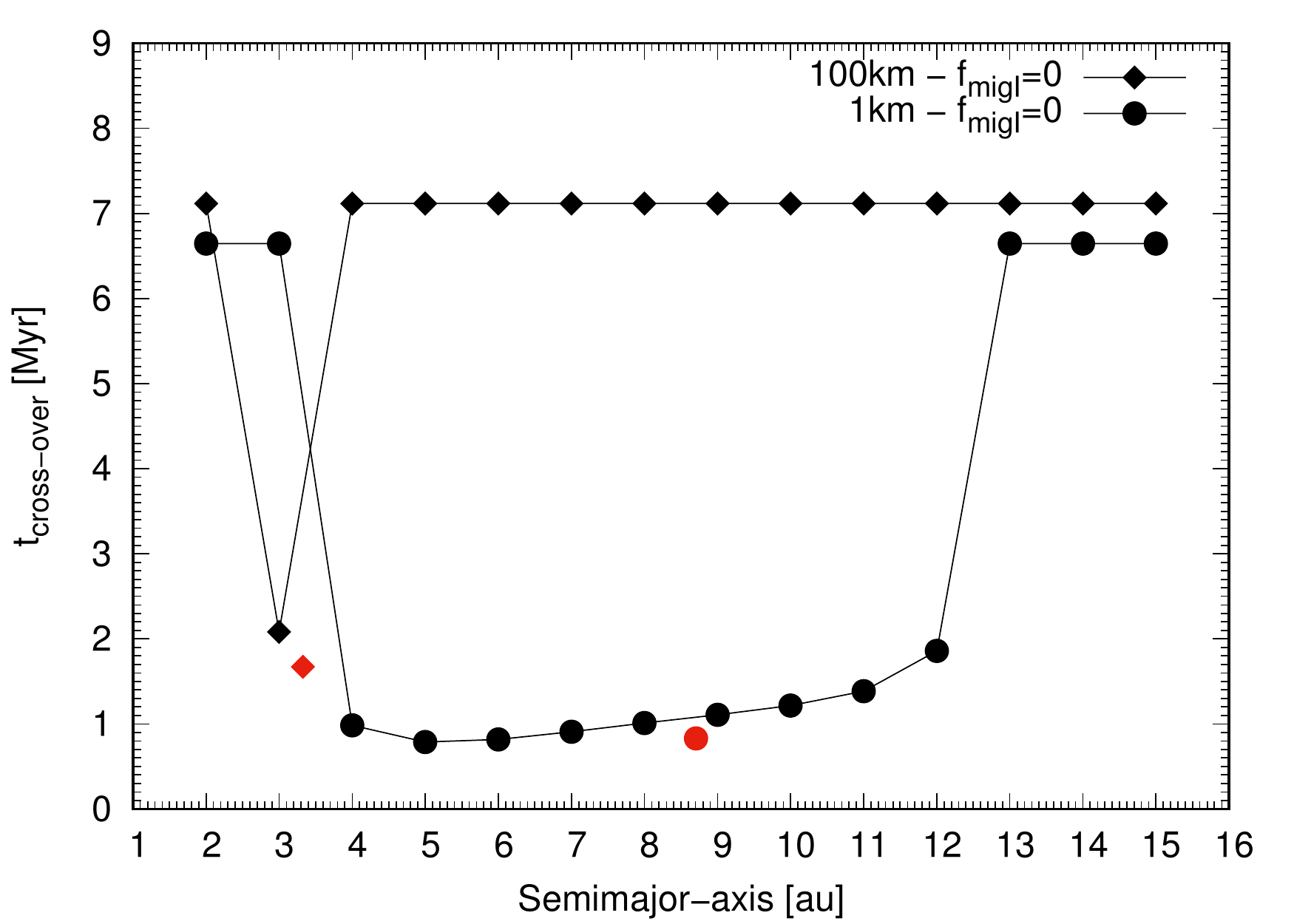}
        \caption{Cross-over time as a function of the initial position of a planet, for the isolated formation, and for scenarios with planetesimals of 1~km and 100~km, and
          without type I migration. The black dots show the isolated formation in scenarios with planetesimals of 1~km. The disc parameters of this formation scenario are
          the same as in panel d) of Figure~\ref{fig:8}, which dissipates in 6.64~Myr. The black diamonds show the isolated formation in scenarios with planetesimals of 100~km, and the disc parameters of this
          formation scenario are the same as in panel i) of Figure~\ref{fig:8}, which dissipates in 7.12~Myr. The red dot and red diamond show the location and time when the first planet of each system (corresponding to
          panels d) and i), respectively), reached the mass of cross-over.} 
    \label{fig:9}
\end{figure}

As an example, Figure~\ref{fig:9} shows the time of cross-over as a function of the position of the planet in the disc. 
The time of cross-over is the time at when
the mass of the envelope of a planet equals the mass of the core (mass of cross-over) and the gaseous runaway triggers. 
The black points and black diamonds in
Figure~\ref{fig:9} represent the in situ formation of an isolated planet, located at different positions between 2~au and 15~au, for the same disc parameters of the
formation scenario of panels d) and i), of Figure~\ref{fig:8}, respectively. As we can see, for scenarios with planetesimals of 1~km, 
the optimal location for the formation
of a giant planet is around 5~au, while for scenarios with planetesimals of 100~km, the optimal location is 3~au. It is worth mentioning that we find similar results
between the scenarios of 1~km and 100~m, and between the scenarios of 10~km and 100~km. However, when we include the formation of the whole complete system, the optimal location
for the formation of the giant can change, and can move away from the central star. As Figure~\ref{fig:9} displays, for scenarios with planetesimals of 1~km the red point in Figure~\ref{fig:9}
shows the location at where the first planet of the system reached the mass of cross-over. This situation shows that, when the simultaneous formation is considered,
the formation of giant planets can be optimized in outermost regions. This is due to the multiple mergers between embryos that have taken place during the formation of the
system. For planetesimals of 100~km, the location of the giant planet is optimized a little bit away from the position found for the isolated formation. In this scenario,
as the embryos do not grow enough to increase their Hill radius, the rate of mergers is lower.

Continuing with the description of Figure~\ref{fig:8}, if we only consider those planetary systems without type I migration, to distinguish the effect caused by the increase
in the planetesimal size, (see panels a, d, f and i), we can see that for 
planetesimals of 100~m, the giant planets form in the outer regions of the disc. Since the accretion rates are higher than for big 
planetesimals, massive cores form in a shorter timescale, and thus, the planets in the outer region can merge between them before the 
dissipation of the gas disc, giving rise to giant planets in the outer zones. These
giants clean their surroundings of planetesimals, leaving a very small remnant almost beyond 20~au. Despite both planets are 
Jupiter-analogs, the innermost giant opened a gap in the disc and moved inward through type II migration, while the outermost did not. 
A brief discussion about why a giant planet with $\sim 4.8\text{M}_{\text{J}}$ did not opened a gap in the disc is discussed in the next section.

Following with the scenario of planetesimals of 1~km (panel d), as the planetesimal accretion rates decrease with respect to the 
previous scenario, those planets in the outer zone grow achieving masses between $\sim 1.5\text{M}_\oplus$ and $\sim 6.5\text{M}_\oplus$. 
These planets do not grow enough to merge between them and thus, we do not find giants on the outer part of the disc. Moreover, they do not
remove their feeding zones completely and do not empty them from planetesimals. Therefore, at the end of the gas phase, a population of 
planets of several Earth masses and a population of planetesimals coexist in the outermost regions of the disc.
A similar situation occurs for scenarios with planetesimals of 10~km (panel f). But in this case, the optimal location for the formation of the giants moves
inward a little bit. At last, for scenarios with planetesimals of 100~km, the formation of giant planets is optimized near the snowline, at $\sim 3$~au. In panel
i), the only formed giant reached 2~au due to the fact that opens a gap in the disc and moves through type II migration.

Regarding the population of planetesimals, a remnant of planetesimals survived at the end of the gaseous phase in the majority of the performed simulations inside 30~au.
However, it is worth mentioning that we found a few simulations of planetary systems without planetesimals inside 30~au, for scenarios with planetesimals of 100~m. How this result will affect, or not, the post-oligarchic growth of the planets in these system is an analysis that we are going to take into account in our next work.

Another important thing to note, with respect to the final planetesimal population, is that some planetary systems, present a small remnant of planetesimals inside the
location of the giant planet. This can be seen, for example, in panels c) and h) of Figure~\ref{fig:8}. Particularly, the remnant of planetesimals in the formation scenario of panel h) is located between 2~au and 3~au. This
seems to be of interest, since it could lead to the formation of an asteroid belt between the terrestrial-planet forming region and the giant planets. However, it will be important to analyze in a future if these planetesimals would survive in the disc, since after the dissipation of the gas they present eccentricities and inclinations greater than $\sim 0.4$ and $10^{\circ}$.

One of the differences between our model model of planet formation and other population synthesis models is that we include a detailed 
treatment of the orbital evolution of the planetesimal population. This improvement allows us to obtain more realistic planetesimal surface 
density, eccentricity and inclination profiles at the end of the gaseous phase to use as initial conditions for the planetesimal population 
in the post-gas evolution of these systems. 

\subsubsection{Gap opening planets}
\label{subsubsec:subsubsec20-0}

\begin{figure}
	\includegraphics[angle=270, width=0.95\columnwidth]{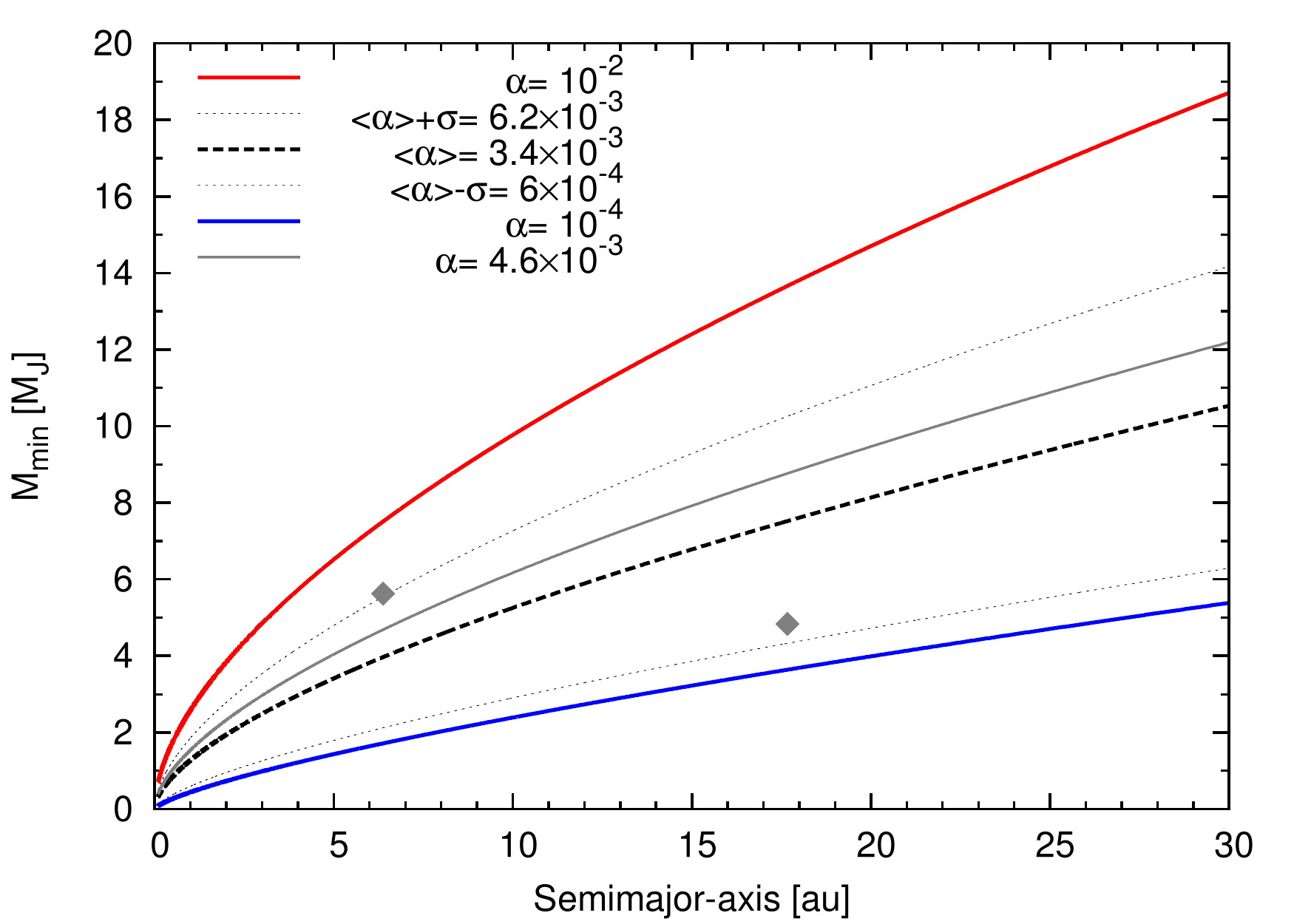}
        \caption{Minimum mass needed to open a gap as a function of the planet semimajor-axis according to \citet{Crida2006}. The red curve corresponds to a disc with $\alpha= 10^{-2}$ (the upper-limit case), while the blue curve corresponds to a disc with $\alpha= 10^{-4}$ (the lower-limit case). The black dashed line represents a disc with mean value of $\alpha$ (for the SSA), $<\alpha>= 3.4 \times 10^{-3}$, and the black dotted lines represent discs with $<\alpha> \pm \sigma$. Finally, the grey curve represents the system described in panel (a) of Fig.~\ref{fig:8} The grey diamonds represent the final mass and final semimajor-axis of the giant planets of this system.}
    \label{fig:10}
\end{figure}

In SSA the frequency of a gap opening planet in the disc is relatively low. For each forming-giant planet that opens a gap in the disc, there are 5.5 giant planets that did not. As we described in Sec. \ref{subsubsec:subsubsec10}, the condition that a giant planet must fulfill to open a gap, is that proposed by \citet{Crida2006}. This condition depends basically on the mass and the location of the planet, and on the viscosity and the height scale of the disc. More explicitly, this condition depends on the sum of two terms: the first term involves the ratio between the height scale of the disc and the Hill radius of the planet, this last is a function of the mass and position of the planet; the second term involves the ratio between the viscosity of the disc and a function of the location and mass of the planet. Thus, the opening of a gap is favored by massive planets, and low viscosity and flat/thin discs (i.e. discs with small aspect ratios $H_{\text{g}}/R$). However, our model considers a flare disc, hence, $H_{\text{g}}/R \propto R$ and it increases with the distance to the central star. This situation coupled with having moderate/high values of the viscosity of the discs, results in the low occurrence of gap opening for giant planets in our model. As an example, Figure~\ref{fig:10} shows the minimum mass needed to open a gap, adopting the criteria developed by \citet{Crida2006}, as a function of the planet semimajor-axis for discs with different viscosities. We show the limit cases for the values of the $\alpha$ parameter adopted in this work, and the mean value ($\pm$ the dispersion) of such parameter. We also plotted the curve corresponding to the planetary system described in panel (a) of Fig.~\ref{fig:8} (grey line), and also the final location of both giant planets of that system.  We can see that the inner Jupiter-analog has a final mass greater than the minimum needed to open a gap in such disc. However, the outer Jupiter-analog, despite of having a mass of $\sim 4.8~\text{M}_{\text{J}}$, did never achieved the minimum mass needed to open a gap.

\subsubsection{Water accretion on planets of the habitable zone}
\label{subsubsec:subsubsec20}

A point of interest for our SSA is to analyze the final amounts of water by mass in those planets that remain, at the end of the gaseous phase, in the habitable zone.
Taking into account the classical definition, the habitable zone is defined as the range of heliocentric distances at which a planet can retain liquid water
on its surface \citep{Kasting1993}. Ten years later, \citet{Kopparapu2013,Kopparapuerrata2013} proposed that our solar system presents 
 a conservative habitable zone determined by loss of water and by the maximum greenhouse effect provided by a CO$_{2}$ atmosphere, between 0.99~au and 1.67~au. Moreover, these authors proposed that there is also an optimistic habitable zone between 0.75~au and 1.77~au, limits that were estimated based on the belief that Venus has not had liquid water on its surface for the last 1 billion years, and that Mars was warm enough to maintain liquid water on its surface. Although the definition of the habitable zone is a topic under
permanent debate, and although there is no guarantee that a planet inside this region will be potentially habitable, we consider this definition just as a first
approximation. As it was mentioned before in Sec. \ref{subsec:subsec11}, both embryos and planetesimals present, at the beginning of our simulations, a water radial distribution. As
the embryos grow and the planetesimal surface density evolves, also the amount of water on each body changes. It is important to notice that, as we can see in Figure~\ref{fig:8}
those planets in SSA that remain within the habitable zone at the end of the gaseous phase do not present any water contents. This is due to the fact that the embryo distribution beyond the snowline act as a barrier preventing that water rich planetesimals from the outer zone are 
accreted by embryos of the inner zone. Moreover, the low/null type I migration rates also prevent that embryos from beyond 2.7~au reach
the habitable zone. Although these planets are dry when the gas in the disc dissipates, this does not
mean that they can not acquire significant amounts of water for the possible development of life during the post-oligarchic growth stage via, for example, accretion of wet embryos or planetesimals from beyond the snowline. This analysis is out of the scope of this work but is going to be taken into account in PII.


\section{Summary and discussion}
\label{sec:sec21}

During the last years, advances in observational astronomy have allowed us to increase the sample of confirmed exoplanets very quickly.
This sample, which continues to grow day by day, has motivated the theoretical astronomers to improve their models of planet formation
to reproduce the most important properties of this population. 

In this first work, we improved our model of planet formation, P{\scriptsize LANETA}LP, incorporating relevant physical phenomena to the
formation during the gaseous phase. In our model, the gaseous component of the disc evolves as a viscous accretion disc with photoevaporation 
due to the central star. We adopted an EUV-photoevaporation model based in the works of \citet{Dullemond2007} and \citet{DangeloMarzari2012}.
Many theoretical studies link the photoevaporation process with the observed inner holes in transition discs \citep{OwenClarke2012,Owen2016,Terquem2017}.
Particularly, \citet{Ercolano2017}, were able to show that the gap in the disc around TW-Hya, which is the closest protoplanetary disc to the Earth,
is consistent with their photoevaporation models. However it is important to note that these models are a combination of EUV and X-rays (or just X-rays)
as photoevaporation sources. The percentage of observed transitional discs, which are discs with an inner hole of several au, respect to the total
number of observed discs, is relatively small, of about $10\%-20\%$. The low percentage of observed transition discs suggests that once the gap is
opened in the disc by photoevaporation, the disc should dissipate in a $10\%-20\%$ of the dissipation timescale. In this sence, it is important to note that
our model of purely EUV-photoevaporation source could be overestimating the dissipation time after the gap is opened.

The model also presents a solid component, composed by a planetesimal and embryo population.
The code calculates in detail the orbital migration of the planetesimal population due to the gas drag, and the evolution of their 
eccentricities and inclinations due to the embryo gravitational excitation, and damping by the gas. The embryos grow in the disc due to 
the accretion of planetesimals and the surrounding gas. P{\scriptsize LANETA}LP also considers the fusion between embryos, and planets with 
significant atmospheres. It is worth mentioning that the gas accretion of the planets is not calculated by solving the equations of 
transport and structure for the envelope. Instead, we fit the results of the giant planet formation model
of \citet{Guilera2010,Guilera2014}. Type I and type II migration are also included for the embryos. We also incorporate a water 
radial distribution for both populations.

The principal aim of this work, is to find scenarios and disc parameters that favor the formation of solar system analogs. Obtaining final
embryo distributions and final planetesimal density, eccentricity and inclination profiles of these solar system analogs at the end of the 
gas phase will allow us to study and analyze in a future work (PII), the post-oligarchic growth of these 
systems, using N-body simulations. 

First, we define initial conditions for our simulations, considering that we are interested 
in generating a great diversity of planetary systems without following observable distributions for the disc parameters. However, the 
bounds of the disc parameter ranges considered, come from previous observational works \citep{Andrews2010}. 
We also define different scenarios for the type I migration, introducing a reduction factor $f_{\text{migI}}$, which takes into account 
that the planet migration could be not considered ($f_{\text{migI}} = 0$), slowed down 10 times ($f_{\text{migI}} = 0.1$), 100 times ($f_{\text{migI}} = 0.01$),
or not slowed down at all  ($f_{\text{migI}} = 1$). Besides, we consider different planetesimal sizes of 100~m, 1~km, 10~km and 100~km.
In order to find gas discs that dissipate in timescales according to the observed ones, we run a version
of P{\scriptsize LANETA}LP limited to the study of the evolution of the gas disc, without considering the evolution
of the embryo and planetesimal population. We then generate as many gas discs with dissipation
timescales between 1 and 12~Myr as we need. Finally, we automate our planet formation model to generate a population synthesis of 
16000 planetary system simulations, separated in 16 blocks of different planetesimal sizes and type I migration rates.  

Our general results show that the most common planetary systems are those
with only rocky planets. Moreover, these systems represent more than $60\%$ of the total of the performed simulations. This general result is in
agreement with previous works of population synthesis analysis \citep{Thommes2008,Mordasini2009,Miguel2011}. Distinguishing the formation scenarios,
rocky planetary systems are majority in scenarios with big planetesimals and low-moderate/null type I migration rates. On the contrary, icy giants
are mostly in scenarios with small planetesimals and high type I migration rates. Finally, to most suitable formation scenarios to form giant planets
are those with small planetesimals and low-moderate/null type I migration rates. Those giants formed with high migration rates end up as hot-Jupiters,
very near the central star.

Regarding solar system analogs (SSA), classified as those planetary systems with only rocky planets in the inner zone of the disc and with at least one giant planet
beyond 1.5~au, they represent only a $4.3\%$ of the total of performed simulations. The most representative SSA architectures found are those that present
between 1 to 3 gas giants, between 0 to 4 icy giants, and between 100 to 200 rocky planets along the disc. The most favorable formation scenarios for this kind of systems are those
with small planetesimals and low/null type I migration rates. SSA were preferentially formed in discs with
smooth surface densities profiles ($<\gamma>= 0.95$), moderate values for the characteristic radius ($<R_{\text{c}}> \sim 33$~au), small values of the $\alpha$-viscosity
parameter ($<\alpha>= 3.4 \times 10^{-4}$), moderate timescales for the disc dissipation ($<\tau> \sim 6.5$~Myr) and massive discs ($<M_{\text{d}}> \sim 0.1\text{M}_\odot$). 
Although we found SSA in almost all the disc range parameters considered, we did not find them in low-mass discs, with $M_{\text{d}} < 0.04\text{M}_\odot$. Regarding water delivery, we found that those embryos that remained within the habitable zone, were totally dry when the gas disc had dissipated.

Despite our model of planet formation includes important physical phenomena for the formation of a planetary system during the gas phase, 
it yet does not include many fundamental interactions that may affect our final results. Perhaps, the most important are:
\begin{enumerate}
\item Type I migration rates for non ideal isothermal discs. \citet{Paardekooper2010,Paardekooper2011} found that type I migration rates for 
non isothermal discs can be different with respect to those calcuted previously for isothermal discs. Type I migration rates for isothermal 
discs only depend on the disc surface density profile, while for more realistic discs, these migration rates also depend on the temperature 
profile of the disc. Moreover, while ideal isothermal discs in general lead to a rapid inward migration, type I migration for non isothermal discs can be outward, depending on the temperature structure of the disc, and on the mass and semimajor-axis of the planet. \citet{Dittkrist2014} studied the impact of planet migration models on planetary population synthesis incorporating type I migration rates for non isothermal discs. Comparing population synthesis adopting migration rates for isothermal (without any reduction factor) and non isothermal discs, they found that despite the M-a diagrams are similar for both cases, the percentage of planets that remain outside 0.1~au in the first case is $\sim$ the half, compared to the second case. Thus, type I migration rates inferred for more realistic discs could change the percentage of our SSA. Otherwise, \citet{ColemanNelson2014,ColemanNelson2016} also incoporated type I migration rates for non isothermal discs in a planet formation model. These authors found that the formation of giant planets is favored by the accretion of small planetesimals ($r_{\text{p}} < 100$~m) and that the existence of radial disc structures is necessary for the survival of such planets in the outer parts of the discs. However, we note that the planet formation model of \citet{ColemanNelson2014} is quite different with respect to our own, or the one developed by \citet{Dittkrist2014}, specially in the treatment of the planetesimal accretion and growth of the planets. Finally, it is important for us to remark again that, to be consistent with our assumption of isothermal disc, we followed \citet{Tanaka2002} prescriptions for type I migration rates. The impact of considering more realistic discs and type I migration rates for non isothermal discs in the formation of SSA will be subject of a future work.
\item Gravitational interactions between planets and mean-motion resonances. A key effect, that might significantly alter the results of our simulations, is the fact of considering
  gravitational interactions between the planets. These interactions could cause two different effects. On the one hand, when the gas is almost removed in the inner regions of the disc due to photoevaporation, the dispersion between gaseous giant planets can lead to the ejection of one or two planets exalting the eccentricities of the remaining ones, although there is still gas in the outer zones. Moreover, \citet{Matsumura2013} showed that even the terrestrial planets of the inner zone of the
  disc can be affected by the giants, despite being far from them, in discs without gas. These effects can alter the final configurations of our SSA.
  On the other hand, gravitational interactions could allowed the planets to be trapped in mean motion resonances (MMR). The migration of planets trapped in MMR along the gaseous disc is a complex phenomenon. Towards this mechanism, planets trapped in MMR could be able to avoid a fast orbital decay into very inner zones of the disc \citep{MassetSnellgrove2001}. 
  Moreover, regarding the formation of our solar system, if Jupiter was able to open a gap in the disc and migrate inward through type II migration, and at the same time Saturn was
  able to migrate faster towards Jupiter, they could have been locked in a 2:3 (MMR). This effect could have stopped or even reversed the migration of both planets toghether \citep{MorbidelliCrida2007}. Then, \citet{Morbidelli2007} showed that if the gas giant planets were trapped in a MMR, the icy giant planets could also be trapped in MMR and all the system could evolve locked in a MMR chain. These results are of great importance, since they represent the initial orbital configuration of the outer solar system after the gas disc is dissipated, assumed by the Nice model \citep{Tsiganis2005,Gomes2005,Morbidelli2005}. However, it is important to note that several phenomena like, for example, disc turbulences, could break the MMR configurations \citep{Adams2008}. The inclusion of the treatment of these interactions in P{\scriptsize LANETA}LP is one of our future goals.
\item Planetesimal fragmentation. Planetesimal collisional evolution could have and important impact in the population synthesis results. In fact,
\citet{Guilera2014} developed a planetesimal fragmentation model to study the role of planetesimal fragmentation in giant planet formation. In line with the results 
of the pionner work of \citet{Inaba2003}, and \citet{OrmelKobayashi2012}, the authors also found that substantial amounts of mass could be lost in the planetesimal
collisional evolution process reducing the efficiency of planet formation, specially for small planetesimals, which have a lower specific impact energy.
However, it is important to remark that these works considered that all the mass generated by the planetesimal fragmentation process distributed below some minimum
particle size, is lost. However, \citet{Chambers2014} found that relaxing this hypothesis, considering that the very small particles below such minimum particle 
size can quickly coagulate avoiding the loss of material by the collisional process, planetesimal fragmentation could favour the planet formation process. It is important
to remark that the inclussion of planetesimal fragmentation in a population synthesis analysis is very costly numerically speaking and imposes a very important limitation. 
\end{enumerate}

Finally, it is important to remark that our 
results provide us with planet distributions and planetesimal density, eccentricity and inclination profiles at the end of the gaseous 
phase. These distributions will be use as initial conditions to develop N-body simulations, with the aim of studying the post-oligarchic 
stage of formation, focusing on the formation of terrestrial
planets and their final amounts of water. The planet distributions provide us with information about the location, core and envelope 
mass of the final planets, as well as water contents due
to the planets primordial contents and due to the accretion of embryos and planetesimals. We also obtain detailed characteristics of the 
planetesimal population regarding their final eccentricities,
inclinations and also, water contents. The role and importance that these details may have in the final results and the 
stability and dynamical evolution of these systems in the post-gas phase are subjects that will be developed in the next work (PII).

\section*{Acknowledgements}
We  thank the referee, John Chambers, for constructive comments that helped us to improved the manuscript.
This work is supported by grants from the National Scientific and
Technical Research Council through the PIP 0436/13, and National University of La Plata, Argentina,
through the PIDT 11/G144.




\bibliographystyle{mnras}
\bibliography{Biblio} 








\bsp	
\label{lastpage}
\end{document}